\documentclass[prl,twocolumn,aps,superscriptaddress,longbibliography,10pt]{revtex4-2}
\usepackage[ansinew]{inputenc}
\usepackage[caption=false]{subfig}
\usepackage{bm,color,soul,amsmath,amssymb,mathrsfs,latexsym,graphicx,booktabs,array,psfrag,float,xcolor}

\usepackage[english]{babel}
\DeclareMathOperator{\sgn}{sgn}
\usepackage{verbatim}
\usepackage{enumitem}

\begin{document}

\title{\Large\textbf{Zoology of non-Hermitian spectra and their graph topology}}
\author{Tommy Tai}
\affiliation{Cavendish Laboratory, University of Cambridge, JJ Thomson Ave, Cambridge CB3 0HE, UK}
\email{ytt26@cam.ac.uk}
\author{Ching Hua Lee}
\affiliation{Department of Physics, National University of Singapore, Singapore 117542}
\email{phylch@nus.edu.sg}
\date{\today}
\def\blue{\textcolor{blue}} 
\setlength{\parindent}{0.3cm}
\begin{abstract}
We uncover the very rich graph topology of generic bounded non-Hermitian spectra, distinct from the topology of conventional band invariants and complex spectral winding. The graph configuration of complex spectra are characterized by the algebraic structures of their corresponding energy dispersions, drawing new intimate links between combinatorial graph theory, algebraic geometry and non-Hermitian band topology. Spectral graphs that are conformally related belong to the same equivalence class, and are characterized by emergent symmetries not necessarily present in the physical Hamiltonian. The simplest class encompasses well-known examples such as the Hatano-Nelson and non-Hermitian SSH models, while more sophisticated classes represent novel multi-component models with interesting spectral graphs resembling stars, flowers, and insects. With recent rapid advancements in metamaterials, ultracold atomic lattices and quantum circuits, it is now feasible to not only experimentally realize such esoteric spectra, but also investigate the non-Hermitian flat bands and anomalous responses straddling transitions between different spectral graph topologies.
\end{abstract}
\maketitle

\noindent{\textit{Introduction.--}} Topological classification plays an indispensable role in modern condensed matter physics, identifying the intrinsic robustness in ionic compounds~\cite{schindler2018HOTI,PhysRevX.11.041064} and engineered metamaterial platforms like photonic~\cite{PhysRevLett.100.013904,wang2009observation,lu2013weyl,lu2014photonics,lin2017line,PhysRevB.98.205147,PhysRevX.6.021007,klembt2018exciton,song2019breakup,zhu2020photonic}, mechanical~\cite{kane2014topological,peano2015topological,susstrunk2015observation,nash2015topological,ghatak2020observation,PhysRevLett.125.118001,PhysRevLett.124.073603} and electrical setups~\cite{ningyuan2015time,imhof2018topolectrical,hofmann2019chiral,wang2020circuit,li2019emergence,olekhno2020topological,helbig2020generalized,zou2021observation,stegmaier2021topological,lenggenhager2021electric,yang2021observation}.

Conventionally, topological classification pertains to the classification of eigenstate windings, specifically the topology of the mapping between the Brillouin zone (BZ) and and the target state space. This notion of topology underscores all topological insulators~\cite{kane2005quantum,kane2005z,konig2007quantum,fu2007topological,fu2007,moore2009topological,gu2016holographic,hasan2010colloquium}, higher-order topological insulators~\cite{Schindlereaat0346,benalcazar2017HOTI,song2017HOTI,Fulga2018,edvardsson2019non} and indeed practically all Hermitian topological lattices, symmetry-protected or otherwise. The accompanying topological invariant is connected to a quantized observable such as Hall conductivity~\cite{halperin1982quantized,macdonald1984quantized,haldane1988model,kane2005quantum,kane2005z,konig2007quantum}~\footnote{But see~\cite{li2021quantized}, which concerns a quantized measurable quantity for winding in the energy plane.}, which is as such protected from continuously degrading. 

\begin{figure}
\includegraphics[width=0.95\linewidth]{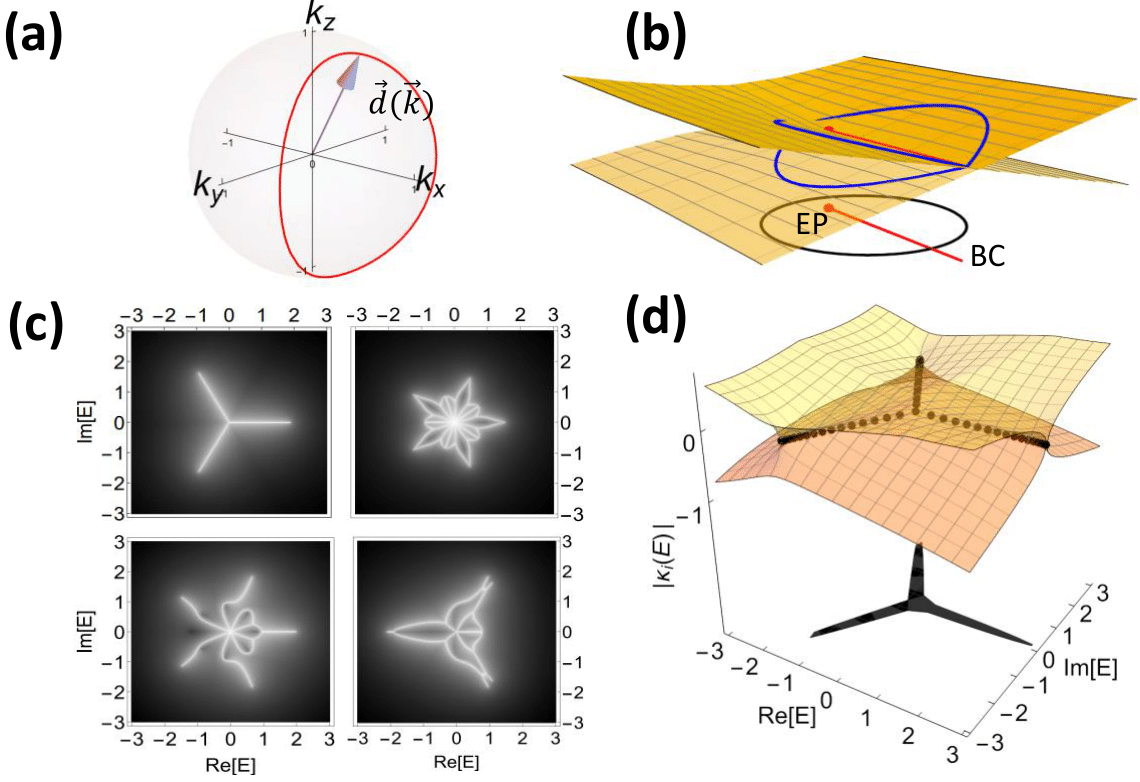}
\caption{Conventionally, topology typically refers to the winding number either (a) in the state space or (b) around the exceptional point (EP) branch cut (BC). (c) This work uncovers the very intricate \emph{graph} topology of the spectra of non-Hermitian bounded lattices, which originates from (d) the intersections of its inverse skin depth solution surfaces $|\kappa_i(E)|$.
}
\label{fig:1}
\end{figure}

In this work, we focus on a different type of topology, namely the \emph{spectral graph} topology, that is found to be far more intricate and exotic than conventional $\mathbb{Z}$ or $\mathbb{Z}_2$ topological~\cite{kane2005z,liu2016quantum} classes. As presented in Fig~\ref{fig:1}c, the energy spectra of various bounded non-Hermitian lattices take on a kaleidoscope of interestingly shapes resembling stars, flowers or even insects,  consisting of lines or curves that connect spectral vertices in all imaginable ways. Compared to eigenstate (Fig.~\ref{fig:1}a) or exceptional point (Fig.~\ref{fig:1}b)~\cite{heiss2012EP,zhen2015spawning,Hassan2017EP,hodaei2017enhanced,Hu2017EP,gao2018chiral,zhang2018dynamically,zhou2018observation,Yoshida2019ER_2,jin2019hybrid,dora2019kibble,denner2021exceptional,lee2022exceptional} topology which are represented by homotopy windings, the planar graph topology of these non-Hermitian spectra can be much more sophisticated, encoding arbitrarily complicated connectivity structures. Indeed, the number of distinct planar graphs with $\mathcal{N}$ branching vertices scales rapidly as~\cite{gimenez2009asymptotic} $\sim \mathcal{N}^{-7/2}\gamma^\mathcal{N}\mathcal{N}!$ with $\gamma\approx 27.27$, and no single topological invariant can unambiguously distinguish between two different graphs.

Just like how conventional eigenstate topology manifests as linear response quantization, topological transitions between different spectral graphs physically manifest as linear response \emph{kinks}. This is because different parts of the eigenstates get to mix abruptly when at transitions between different graph configurations, resulting in enigmatic \emph{gapped} marginal transitions with no Hermitian analog~\cite{li2020critical}. In the simplest cases, such transitions have also been associated with Berry curvature discontinuities~\cite{lee2020unraveling}. Since real experimental setups are almost always finite and bounded, the requirement of open boundary conditions (OBCs) do not diminish the physical significance of spectral graph topology.

Deep mathematical relationships exist between the spectral graph topology of a system and the algebro-geometric properties of its energy-momentum dispersion. As elaborated shortly, the dispersion can be written as a bivariate Laurent polynomial $P(E,z)$, where $E$ and $z=e^{ik}$ are the energy and complex exponentiated momentum respectively. In lattices with multiple components (bands) or hoppings, the OBC spectral graph quickly become very intricate~\footnote{The exact OBC spectrum quickly become analytically intractable, since due to the Abel-Ruffini theorem, no generic analytic solution is possible for sufficiently high-order polynomials~\cite{ruffini1813riflessioni}.}, and $P(E,z)$ becomes a ``fingerprint'' of the set of its possible graph topologies. This correspondence survives under conformal transformations in the complex energy, which is a vast group of symmetries tying together otherwise  unrelated Hamiltonians. In charting out the classification table for distinct graph topologies, we also uncover emergent symmetries absent in the original Hamiltonian, leading to alternative avenues for engineering real non-Hermitian spectra beyond PT symmetry~\cite{bender1998real,tomoki2018photonics,feng2017non,el2018non,longhi2020non,stegmaier2021topological,long2021real,yang2022designing}

\begin{figure}
\centering
\includegraphics[width=\linewidth]{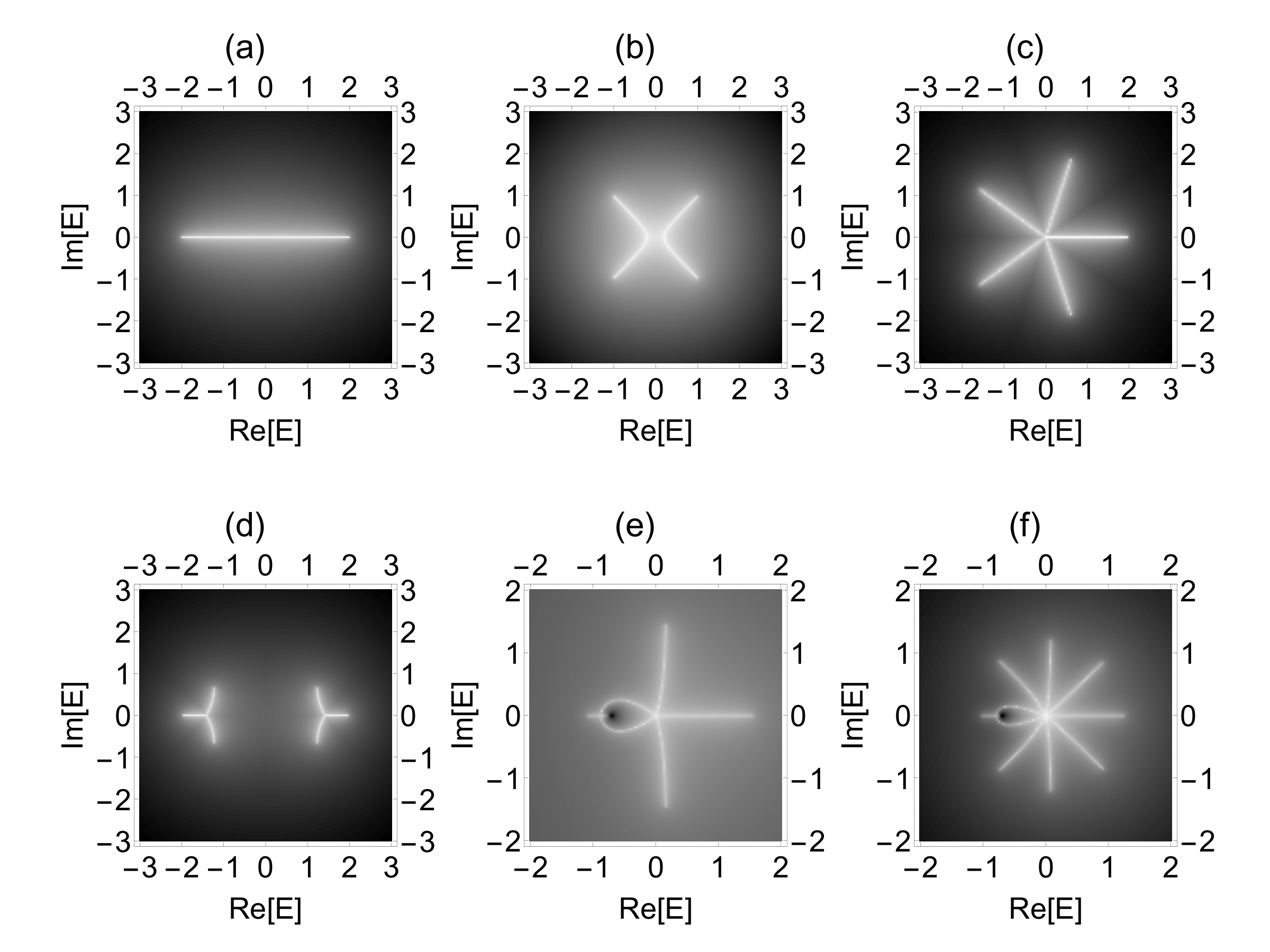}
\caption{
Elementary illustrative OBC spectral graphs $\bar E$: (a) real spectral line segment of the Hatano-Nelson model ; (b) hyperbola spectral segments of the non-Hermitian SSH model with $|\mu|>|t|$; (c) 5-pronged star for $F(E)=E$ and $a=3$, $b=2$; (d) Two deformed 3-stars from the OBC spectrum of $H^\text{2-band}_{2,1}$ with $a=2$, $b=1$ and $F(E)=2-E^2$; (e-f) Looped spectral graphs from Eq.~\ref{real}, with (e) $G(E)=E^2+0.7E$, $F(E)=E^3$ and (f) $G(E)=E^4+0.7E^3$, $F(E)=E^6$.
}
\label{fig:2}
\end{figure}

\noindent{\textit{Spectral graphs from energy dispersions. --}} 
Under OBCs, the spectrum of generic non-Hermitian lattices collapse into straight lines or curves~\cite{lee2019anatomy,yang2019auxiliary,yokomizo2019non} that join to form a planar graph made up of stars and loops [Figs.~\ref{fig:1}c, \ref{fig:2}a-f]. To understand why, we first highlight the fundamental role played by the dispersion relation 
\begin{equation}
P(E,z)=\text{Det}[H(z)-E\, \mathbb{I}]=0,
\end{equation}
which is the characteristic polynomial of the Hamiltonian $H(z)$, $z=e^{ik}$. In an unbounded periodic crystal, the spectrum is simply the set of $E$ satisfying $P(E,e^{ik})=0$ for real momenta $k\in [0,2\pi)$. However, under OBCs, this is not the case since $k$ no longer indexes the eigenstates due to broken translation symmetry. Yet, because the bulk is still translation invariant, any eigenstate must be composed of eigensolutions of Bloch-like form, characterized by \emph{complex} instead of real momenta $k$. The imaginary part of the momentum $\text{Im}(k)$ represents spatial decay rate (since $|e^{ikx}|\sim e^{-(\text{Im}k)x}$), and is also known as the inverse skin depth~
~\cite{Lee2016nonH,gong2018topological,yao2018edge,lee2019anatomy,yang2019auxiliary,yokomizo2019non,yokomizo2019non,lee2020unraveling,schomerus2020nonreciprocal,borgnia2020nonH,zhang2020correspondence,okuma2020topological,kawabata2020higher,xiao2020non,longhi2020non,xue2020non,mu2020emergent,rosa2020dynamics,lee2020ultrafast,lee2020many,li2021impurity,shen2021non,long2021real,jiang2022filling,zhang2021tidal}

In particular, for an eigenstate to satisfy OBCs, it must simultaneously vanish at both ends, and that requires it to be a superposition of degenerate eigensolutions with equal skin depths~\cite{suppmat}\footnote{If the decay rates are not equal, we are left with effectively one eigenstate in the thermodynamic limit, and the state wavefunction cannot vanish at both ends and satisfy OBCs.}. As a consequence, the OBC spectrum consists~\footnote{with the exception of a small number of isolated states protected by eigenstate topology, if any.} of the set of energies $E\in \{\bar E\}$ that satisfy $P(\bar E,z)=0$, and which are simultaneously degenerate in both $E$ and $\text{Im}\,k=-\log|z|$.

Interpreted geometrically, these conditions imply that OBC eigenenergies must lie on a planar graph: On the complex $E$ plane, solutions to $P(E,z)=0$ that are degenerate in both $E$ and $\text{Im}\,k=-\log|z|$ can be visualized geometrically as the intersections of  $\text{Im}\,k$ surfaces. Intersections of two $\text{Im}\,k$ surfaces trace out curves i.e. graph edges, while intersections of three or more $\text{Im}\,k$ surfaces produce branch points i.e. spectral graph vertices [Fig.~\ref{fig:1}d]. Together, these intersection loci trace out a planar spectral graph on the complex plane. Note that if $P(E,z)$ is of degree $p$ in $z$, there exist $p$ surfaces of $\text{Im}\,k$ everywhere due to the fundamental theorem of algebra~\cite{rotharithmetica}.  
Indeed, the OBC spectral graph depends solely on the algebraic form of $P(E,z)$; the exact form of the Hamiltonian $H(z)$ is inconsequential. This is actually already the case in Hermitian lattices - bulk bands are computed from the dispersion relation $P(E,z)=0$, and not explicitly from $H(z)$ per se. What is unique and interesting in the non-Hermitian context is the key role played by $\text{Im}\,k$ solution surfaces across the \emph{entire} complex $E$ plane, not just points satisfying $P(E,z)=0$. 

To understand how $P(E,z)$ determines the possible spectral graphs $\bar E$, we first mention two important symmetries that greatly simply their distinct classification. Firstly, \emph{all} Hamiltonians related by a translation of imaginary momentum $H(k)\rightarrow H(k+ i\kappa)$ i.e. real-space rescaling $c^\dagger_x\rightarrow c^\dagger_x e^{-\kappa x }$ possess~\cite{lee2019anatomy} identical OBC spectra $\bar E $. This is because $\bar E$ depends on the intersections of $\text{Im}\,k$ surfaces, which cannot change if all solution surfaces are translated by an equal amount $\kappa$. As such, all polynomials related by rescalings of $z$ i.e. $P(E,z)\rightarrow P(E,e^{-\kappa}z)$ correspond to the same set of $\bar E$, allowing for the normalization of the coefficients of $z$ (hopping amplitudes) without loss of generality. 

Secondly, different $P(E,z)$ related by a conformal mapping of $E$ possess OBC spectra $\bar E$ related by the same mapping i.e. if $P'(E,z)=P(f(E),z)$, then $\bar E'=f(\bar E)$. In other words, when classifying $P(E,z)$, one only needs to consider the simplest functional dependencies on $E$.

\noindent{\it Elementary examples.-- }
To facilitate our spectral graph classification through $P(E,z)$, we first examine the few simplest examples with analytically known OBC spectra. 

\begin{itemize}[leftmargin=*]
\item \underline{$P(E,z)=F(E)-\left(z+z^{-1}\right)$.}\\
This simplest case can be written in the separable form $F(E)=z+z^{-1}$, where $F(E)$ is solely dependent on $E$, and the right hand side depends only on $z$. It encompasses the two most well-known non-Hermitian lattice models: the Hatano-Nelson and non-Hermitian SSH Models given by~\cite{hatano1996localization,hatano1998non,Lee2016nonH} $H_\text{HN}(z)= uz+vz^{-1}$, and $H_\text{SSH}(z)=(t-\mu+z)\sigma_+ + (t+\mu+z^{-1})\sigma_-$ respectively. For the single-component $H_\text{HN}$, the energy eigenequation is $E-(uz+vz^{-1})=0$, which can be rewritten as $P_\text{HN}(E,z)=E/\sqrt{uv} -(z+z^{-1})=0$ after letting $z\rightarrow \sqrt{\frac{u}{v}}z$ and $F(E)=E_\text{HN}/\sqrt{uv}$. For the 2-component $H_\text{SSH}$, we have $E^2 = (t-\mu+z)(t+\mu+z^{-1})=(t^2-\mu^2+1)+(t+\mu)z+(t-\mu)z^{-1}$, which can also be written as $P_\text{SSH}=F_\text{SSH}(E)-(z+z^{-1})$ upon letting $z\rightarrow z\sqrt{\frac{t+\mu}{t-\mu}}$ and defining $F_\text{SSH}(E)=(E^2+\mu^2-t^2-1)/\sqrt{t^2-\mu^2}$.

Now, since $z+z^{-1}$ is invariant under $z\leftrightarrow z^{-1}$, it is doubly degenerate when $|z|=|e^{ik}|=1$. Thus the OBC spectrum $\bar E$ is just given by $F(\bar E)=2\cos k$, $k\in \mathbb{R}$. For the Hatano-Nelson model, we thus have $\bar E_\text{HN}=2\sqrt{uv}\cos k$, which is a real line interval when $\sqrt{uv}$ is real (Fig.~\ref{fig:2}a). For $H_\text{SSH}$, it is slightly more complicated with $\bar E_\text{SSH} = \pm\sqrt{1+t^2-\mu^2+2\sqrt{t^2-\mu^2}\cos k}$, which constitutes two mirror-imaged real line intervals if $t^2-\mu^2$ is real, and two curved hyperbola segments otherwise (Fig.~\ref{fig:2}b). As such, the spectral graphs $H_\text{HN}$ and $H_\text{SSH}$ both correspond to the same dispersion class $P(E,z)=F(E)-(z+z^{-1})$, and are conformally related to each other.

\item  \underline{$P(E,z)=F(E)-\left(z^a+z^{-b}\right)$.}\\
Next up is the dispersion class $F(E)=z^a+z^{-b}$, which is still separable, but without $z\rightarrow z^{-1}$ symmetry. It occurs when there are dissimilar hopping distances, such as in $H_{a,b}(z)=z^a+z^{-b}$, which contains hoppings of $a/b$ sites to the left/right. A more sophisticated example would be $H_{2,1}^\text{2-band}(z)=(z^{-1}-1)\sigma_++(z+z^2-1)\sigma_-$, which corresponds to $F(E)=2-E^2$ and $a=2,b=1$, and is the 1D precursor to a Chern model without Hermitian limit~\cite{lee2020unraveling}. 

By considering simultaneous rotations of $F\rightarrow Fe^{i\theta}$, $z\rightarrow ze^{i\phi}$, it can be shown that $F(\bar E)$ is a star with $(a+b)$-fold rotational symmetry (Fig.~\ref{fig:2}c). Via a conformal transform, this star may be mapped into multiple distorted stars, such as $\bar E_{2,1}^\text{2-band}$ (Fig.~\ref{fig:2}d).

\item \underline{$P(E,z)=G(E)z+z^{-1}-F(E)$.}\\
This dispersion class is non-separable, containing a mixed term $G(E)z$ involving both $E$ and $z$. It can arise from intra-sublattice hoppings in a multi-component model, such as the minimal model with nontrivial trace
\begin{equation}
H_{\substack{F(E)=E^2,\\ G(E)=E}}(z)=\begin{pmatrix} rz&1/z\\1&0\\\end{pmatrix}.
\end{equation}
In general, mixed terms pose challenges in the analytic characterization of $\bar E$, because they make the GBZ notoriously difficult to express explicitly. However, for this specific ansatz $P(E,z)=G(E)z+z^{-1}-F(E)$, the OBC spectrum is exactly given by~\cite{suppmat}
\begin{equation}
F(\bar E)^2 = \eta\, G(\bar E),\label{real}
\end{equation}
where $\eta\in \mathbb{R}$. This gives star-like spectral graphs when $F(E)$ and $G(E)$ are both monomials, but more esoteric looped topologies otherwise (Fig.~\ref{fig:2}e,f). Note that Hamiltonians corresponding to particular $F(E), G(E)$ are not unique, see \cite{suppmat} for more examples.
\end{itemize}

\begin{table*}
\begin{center}
\begin{tabular}{|c|c|c|c|c|c|c|c|c|c|}
\hline
 & Char. Poly. $P(E,z)$            & $\mathcal{N}_S$ & $\mathcal{N}_L$ & $\mathcal{N}_\ell$  & Symmetry & Minimal Hamiltonian $H(z)$ &    Spectral graph e.g.      \\
\hline
(i) & $z^2+1/z+r~E~z-E^2$           & 3  & 3    & 3                                & 3-fold global, r-reflection
& $\begin{pmatrix}rz& z^2+1/z\\1&0\\\end{pmatrix}$ &  \begin{minipage}{.15\textwidth}
      \includegraphics[width=\linewidth, height=25mm]{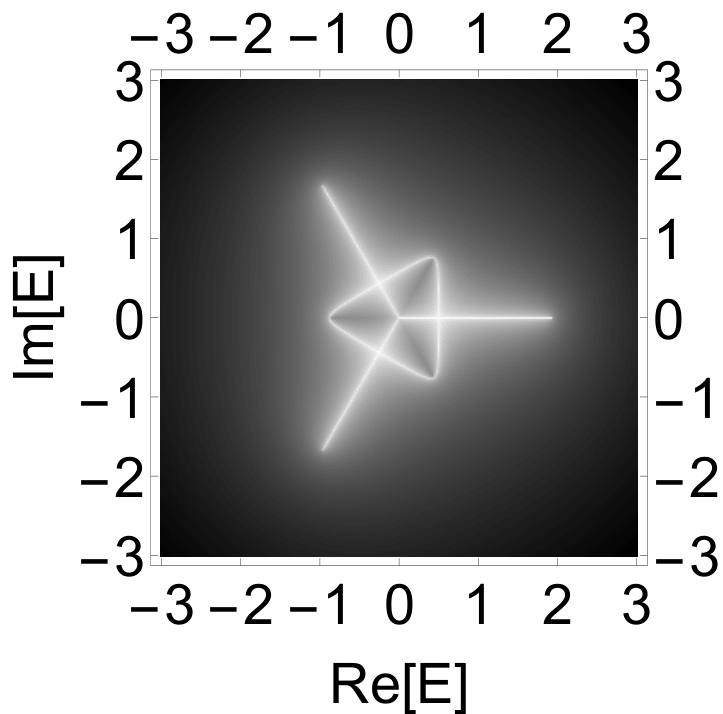}
    \end{minipage}\\
\hline
(ii) &$z^2+1/z+r~E^2~z-E^4$  & 6  & 6  & 6                & 
6-fold global, reflection
 & $\begin{pmatrix}0& rz&0&z^2+1/z\\1&0&0&0\\0&1&0&0\\0&0&1&0\end{pmatrix}$& \begin{minipage}{.15\textwidth}
      \includegraphics[width=\linewidth, height=25mm]{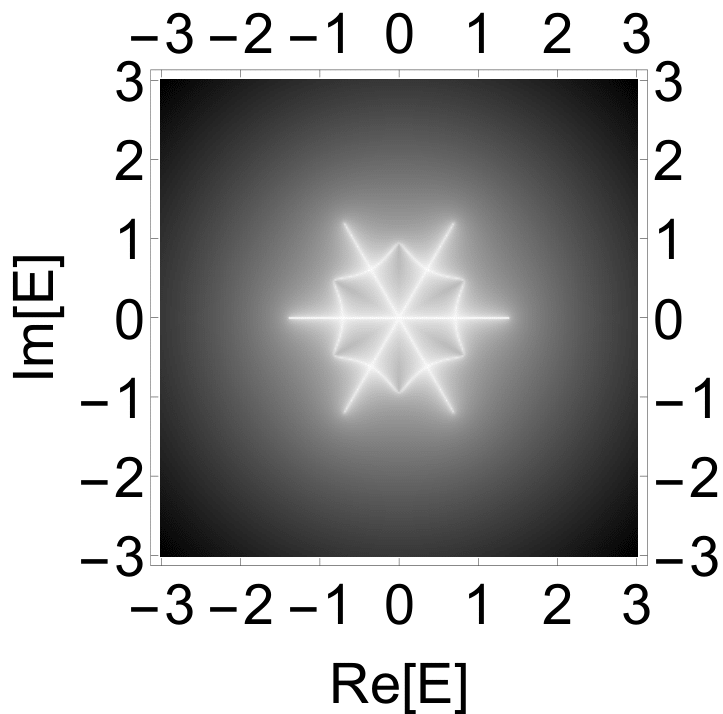}
    \end{minipage}\\
\hline
(iii) &$z^2+1/z+r~E~z-E^3$                    & 3  & 5   & 0              & None                      &$\begin{pmatrix}0& rz&z^2+1/z\\1&0&0\\0&1&0\end{pmatrix}$ &  \begin{minipage}{.15\textwidth}
      \includegraphics[width=\linewidth, height=25mm]{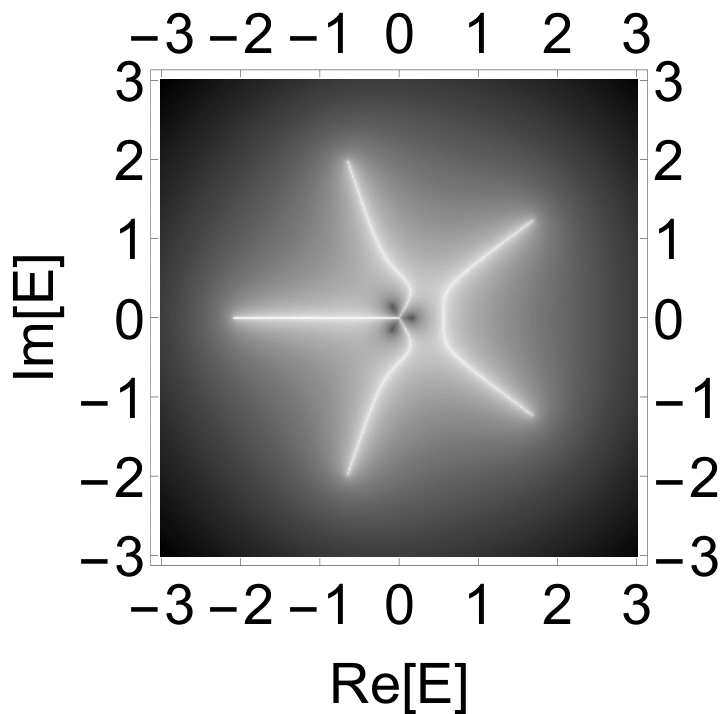}
    \end{minipage}\\
\hline
(iv) &$z^3+1/z^2+r~E~z-E^3$  & 5  & 7   & Depends on $r$                  & None               & $\begin{pmatrix}0& rz&z^3+1/z^2\\1&0&0\\0&1&0\end{pmatrix}$ & \begin{minipage}{.15\textwidth}
      \includegraphics[width=\linewidth, height=25mm]{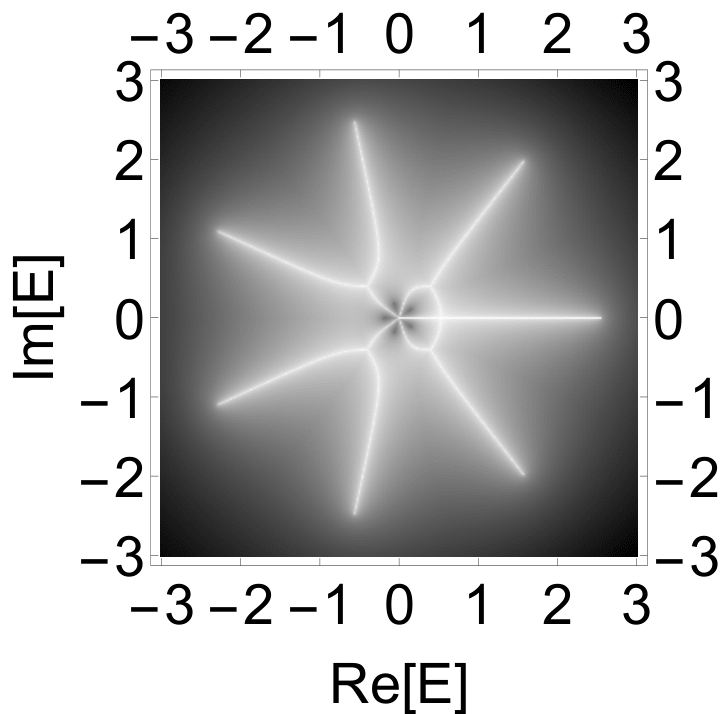}
    \end{minipage} \\
\hline
(v) &$z^2+1/z+r~E^2~z-E^3$   & 6  & 4   & Depends on $r$           & None                     & $\begin{pmatrix}rz& 0&z^2+1/z\\1&0&0\\0&1&0\end{pmatrix}$& \begin{minipage}{.15\textwidth}
      \includegraphics[width=\linewidth, height=25mm]{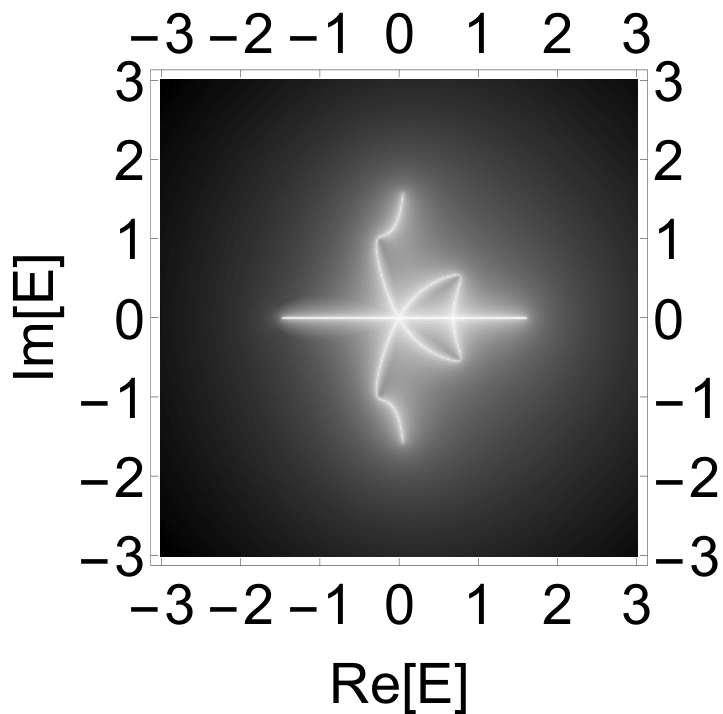}
    \end{minipage} \\
\hline
(vi) &$z^2+1/z+r~E^3~z-E^4$  & 9  & 5   & 3                  & r-reflection               & $\begin{pmatrix}rz& 0&0&z^2+1/z\\1&0&0&0\\0&1&0&0\\0&0&1&0\end{pmatrix}$ & \begin{minipage}{.15\textwidth}
      \includegraphics[width=\linewidth, height=25mm]{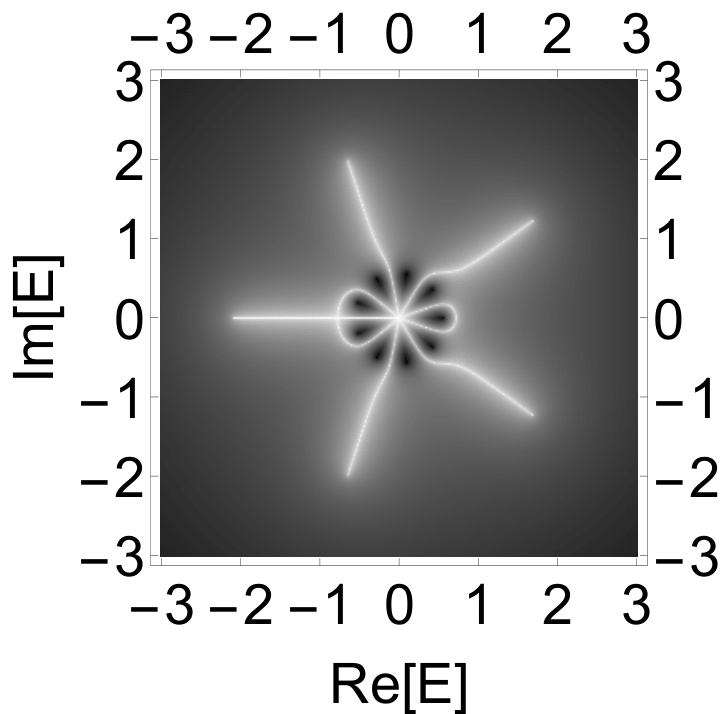}
    \end{minipage} \\
\hline
\end{tabular}
\caption{
Different forms of the canonical dispersion $P(E,z)$ of Eq.~\ref{PEzab} correspond to rich and diverse OBC spectral graphs $\bar E$. For each $P(E,z)$, we can 1. associate a non-unique minimal Hamiltonian $H(z)$, 2. identify emergent global symmetries of $\bar E$ not necessary present in $H(z)$, and 3. characterize its spectral graph topology with its number of branches $\mathcal{N}_S, \mathcal{N}_L$ and loops $\mathcal{N}_\ell$, as well as its adjacency matrix. See the Supplement~\cite{suppmat} for the latter as well as more examples with varying $r$. }
\end{center}
\end{table*}

\noindent{\it Classification of Spectral graphs.--} We are now ready to analyze the spectral graph topology arising from much more general energy dispersions
\begin{equation}
P(E,z)=Q(z)+r\,G(E)J(z)-F(E),
\label{PEzab0}
\end{equation}
which involve not just the sum of $z$ and $E$-dependent terms, but also their product. Representative examples of such $P(E,z)$ are given in Table I. As shown shortly, their OBC spectral graphs depend intimately on their polynomial degrees $f=\text{deg}(F(E))$, $g=\text{deg}(G(E))$, $j=\text{deg}(J(z))$ and $q_\pm=\text{deg}(Q(z^{\pm 1}))$. Notably, $r$, the coefficient for the product term, can drive transitions between graph topologies -- see the final section of the Supplementary~\cite{suppmat} for an extensive array of examples. 
 
While Eq.~\ref{PEzab0} admits no general analytical solution for its $\bar E$, we can comprehensively deduce its spectral graph structure by separately examining its small and large $E$ limits. For definiteness, we specialize to 
\begin{equation}P(E,z)= z^{q_+}+\frac{1}{z^{q_-}}+r~E^g~z^j - E^f,
\label{PEzab}
\end{equation}
which is equivalent to many other realizations of Eq.~\ref{PEzab0} up to conformal transforms of $E$ or rescalings of $z$.

When $E\rightarrow 0$, we have $P(E,z)\sim Q(z)+r\,G(E)J(z)=z^{q_+}+z^{-q_-}+rE^gz^j $, since $g<f=\text{dim}(H)$ from the definition $P(E,z)=\text{Det}[H(z)-\mathbb{I}\,E]$. Clearly, the condition $P(E,z)=0$ is equivalent to that of the previously discussed $P(E,z)=F(E)-(z^{q_+}+z^{-q_-})$, and we conclude that for small $|E|$, the OBC spectrum forms a star with
\begin{equation}
\mathcal{N}_S=f(q_++q_-)
\label{NS0}
\end{equation}
rotationally symmetric branches centered at the origin. Note that since $Q(z)=\text{Det}[H(z)]$, $\mathcal{N}_S$ is limited by $\text{max}(q_+,q_-)\leq \text{dim}(H)\text{range}(H)$, where $\text{range}(H)$ is the maximal hopping distance.

At sufficiently large $|E|$ and $r$, $P(E,z)\sim z^{q_\mp}+rE^gz^j-E^f $, where $\mp=-\text{sgn}(j)$. 
By considering simultaneous phase rotations $E\rightarrow Ee^{i\theta}$, $z\rightarrow ze^{i\phi}$, one finds~\cite{suppmat} that $\bar E$ is $2\pi/\mathcal{N}_L$ rotation-symmetric, where 
\begin{equation}\label{NL0}
    \mathcal{N}_L=\frac{(q_\mp+|j|)f-gq_\mp }{\text{GCD}(q_\mp+|j|,q_\mp)}.
\end{equation}
This emergent combination of $\mathcal{N}_S$ and $\mathcal{N}_L$-fold rotational symmetries 
can be employed for the design of new models with continua of real energy states~\cite{yang2022designing}, even when such symmetry is not obvious from the Hamiltonian.

Armed with Eqs.~\ref{NS0} and \ref{NL0}, one can deduce the topology of the entire spectral graph via the following heuristics: 

\begin{enumerate}[leftmargin=*]
\item At small $|E|$ and large $|E|$, the OBC spectral graph of $P(E,z)$ have respectively $\mathcal{N}_S$ and $\mathcal{N}_L$ rotationally symmetric branches centered at the origin. The rest of the spectral graph interpolates between them.
\item Generally, the origin is the only branching point if $\mathcal{N}_L=\mathcal{N}_S$ (Table I(i)).  An exception might occur if the equal number of spectral branches at small and large $|E|$ are displaced by a small rotation angle. Additional branches appear to connect these two regimes, typically leading to a flower-like shape (Table I(ii)). 
\item When $\mathcal{N}_S<\mathcal{N}_L$, additional disjointed or isolated branches can appear at larger $|E|$ (Table I(iii)). However, this condition alone does not guarantee the existence of isolated branches (Table I(iv) is connected). 

\item Depending on the symmetry of $P(E,z)$ under $r\rightarrow -r$, the spectral graph at sufficiently large $\pm r$ are either identical, or mirror-reflected ($r$-reflection). 

\item As we tune $|r|$, the number of rotationally symmetric branches interpolates between $\mathcal{N}_L$ at large $|r|$ and $f(q_++q_-)$ at small $|r|$, as exemplified in~\cite{suppmat}.

\item Sometimes, branches emanating from the origin may join up into loops. This is especially common when $\mathcal{N}_S>\mathcal{N}_L$ (Table I(v) and I(vi)), but may also appear when $\mathcal{N}_S<\mathcal{N}_L$ (Table I(iv)).
\end{enumerate}
The total number of loops $\mathcal{N}_\ell$ is related to the number of vertices $\mathcal{V}$, branch segments $\mathcal{E}$ and disconnected spectral graph components $\mathcal{C}$ via Euler's formula on a planar graph~\cite{euler1758elementa}: $\mathcal{N}_\ell=  \mathcal{C}+\mathcal{E}-\mathcal{V}.$ 
Here $\mathcal{V}$ includes branching points as well as endpoints of branches - there are usually $\mathcal{N}_L$ of them, as showcased at the end of the Supplement~\cite{suppmat}. 
Consider Table I(iv), we have $\mathcal{V}=13$, $\mathcal{E}=14$ and $\mathcal{C}=1$, and thus $\mathcal{N}_\ell=1+13-11=2$ loops.

All in all, the graph structure is revealed through two different interpolations - that between small and large $|E|$ reveals the branching pattern for a particular $P(E,z)$ at large $|r|$, and further interpolating to $r=0$ reveals additional topological transitions. 
For each graph, one may encode its topology by labeling its branching points and constructing the adjacency matrix~\cite{suppmat}.

\noindent{\it Discussion.--} With much of physics controlled by the competition between energetics and entropy, proper understanding of the energy spectrum structure is central in explaining key optical, thermodynamic and electronic properties~\cite{zanatta2019revisiting,chaves2020bandgap,bernardi2017optical}. In this work, we have uncovered that non-Hermitian spectra can present far richer graph topologies than hitherto reported, with their structure heuristically deducible from the dispersion equation $P(E,z)$ (Table I). The simplest topologies i.e. the $P(E,z)=F(E)-(z+z^{-1})$ class have already been experimentally measured in ultracold atomic lattices~\cite{PhysRevLett.124.070402,PhysRevB.104.155141,liang2022observation} as well as photonic, mechanical and electrical networks~\cite{helbig2020generalized,weidemann2020topological,ghatak2020observation,xiao2020non}. Some of these setups harbor the versatility to realize more sophisticated topologies through larger unit cells and longer-ranged couplings. 

Mathematically, our characterization unveils new symmetries not present in the original Hamiltonian~\cite{Gong2018nonHclass,kawabata2019symmetry,okuma2020topological}, and suggests new links between combinatorial graph theory, algebraic geometry~\footnote{Our results are totally unrelated to known correspondences between polynomials and graphs i.e. chromatic polynomials and Dessin d'enfants~\cite{jones1995dessins,read1968introduction}.} and non-Hermitian band topology~\cite{PhysRevLett.125.118001,PhysRevB.103.205205,li2021non,zhang2021tidal}. One open mathematical question remains: For a given spectral graph topology, can one always find a local parent Hamiltonian~\footnote{A general construction exists by means of an electrostatic mapping~\cite{yang2022designing}, but locality is not guaranteed.}?

Moving forward, exciting discoveries are expected in the following two directions: (1) Multi-dimensional generalizations, since the interplay with hybrid higher-order topology can lead to the spontaneously breaking of symmetries in the spectral graphs, as the $\kappa$ deformation affects topological and bulk modes differently~\cite{lee2019hybrid}. (2) Introduction of interactions, which causes emergent OBC flat bands from spectral graph transitions to become highly susceptible to new many-body effects that are potentially accessible in ultracold atomic setups~\cite{Bernien2017,li2020topological,liang2022observation} and, more recently, quantum circuits~\cite{choo2018measurement, smith2019crossing, azses2020identification, mei2020digital,koh2021stabilizing} with monitored non-unitary measurements~\cite{dogra2021quantum,fleckenstein2022non}.

\section{Acknowledgements}
TT is supported by the NSS scholarship by the Agency for Science, Technology and Research (A*STAR), Singapore.
This work is supported by the Singapore Ministry of Education (MOE) Tier 1 grant (21-0048-A0001).

\bibliographystyle{ieeetr}
\bibliography{references}

\clearpage

\onecolumngrid
\begin{center}
\textbf{\large Supplementary Materials}\end{center}
\setcounter{equation}{0}
\setcounter{figure}{0}
\renewcommand{\theequation}{S\arabic{equation}}
\renewcommand{\thefigure}{S\arabic{figure}}

In these supplementary sections, we aim to provide a self-contained detailed mathematical treatment of our classification results. It can be said that our general motivation here is to understand how to design and obtain open boundary condition (OBC) spectra that are unprecedentedly beautiful and intricate, for instance like those in Fig.~\ref{fig:2} of the main text.

Main sections:
\begin{enumerate}
\item Primer on the determination of non-Hermitian OBC spectra
\item Separable $P(E,z)$
\item Non-separable $(E,z)$ with detailed analysis of symmetries and spectral graphs
\item Adjacency matrices of illustrative spectral graphs
\item Compendium of exotic spectra
\end{enumerate}

\section{Primer on the determination of non-Hermitian OBC spectra}
\label{app:NHSE}
First, let's recall how the eigenenergy spectrum of a Hamiltonian $H$ is determined in general. Under periodic boundary conditions (PBCs), there is translational symmetry, and the eigensolutions are indexed by a momentum index $k$. For a 1D system with $L$ lattice sites, the PBC energy spectrum is simply given by the $L$ values of $E$ that satisfies
\begin{equation}
P(E,z)=\text{Det}[H(z)-E\,\mathbb{I}]=0
\label{CP}
\end{equation}
where $z=e^{ik}$, $k\in \{0,\frac{2\pi}{L},\frac{4\pi}{L},...\}$ and $P(E,z)$ is the characteristic polynomial. However, under open boundary conditions (OBCs), translational invariance is broken, and Bloch states indexed by real momenta $k$ are no longer eigenstates. Instead, in order to simultaneously satisfy the OBCs (vanishing of wavefunctions) at both ends, the OBC eigenstate should be a superposition of two or more degenerate generalized Bloch solutions that also decay equally fast. In other words, there should be at least two complex-deformed Bloch solutions of the same $\text{Im}(k)$ but different $\text{Re}(k)$ at a particular point on the OBC spectrum. Mathematically, the OBC spectrum (which we denote as the set of $\bar E$ points) is thus contained in the set of values of $E$ which satisfies Eq.~\ref{CP} for at least two different $z$ with the same $|z|=e^{-\text{Im}(k)}$. Since $\text{Im}(k)$ determines the decay rate of the eigensolution via $\sim e^{-|\text{Im}(k)|x}$, the $\text{Im}(k)$ solutions should also be the smallest i.e. slowest decaying for a particular $\text{Re}(k)$. This solution to $\text{Im}(k)$ in terms of $\text{Re}(k)$ defines the generalized Brillouin zone (GBZ)~\cite{yang2019auxiliary,song2019realspace,yao2018edge}.
$\kappa=\text{Im}(k)$ is also called the inverse decay length or inverse skin depth.

\begin{figure}
\begin{minipage}{\linewidth}
\subfloat[]{\includegraphics[width=.5\linewidth]{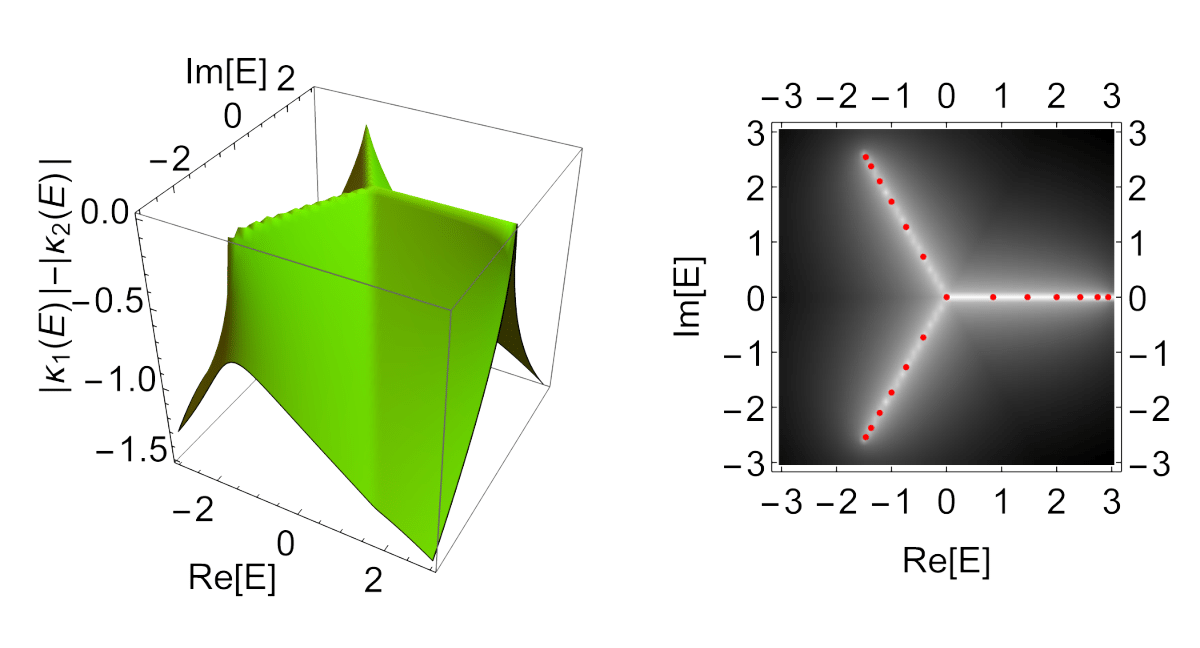}}
\subfloat[]{\includegraphics[width=.5\linewidth]{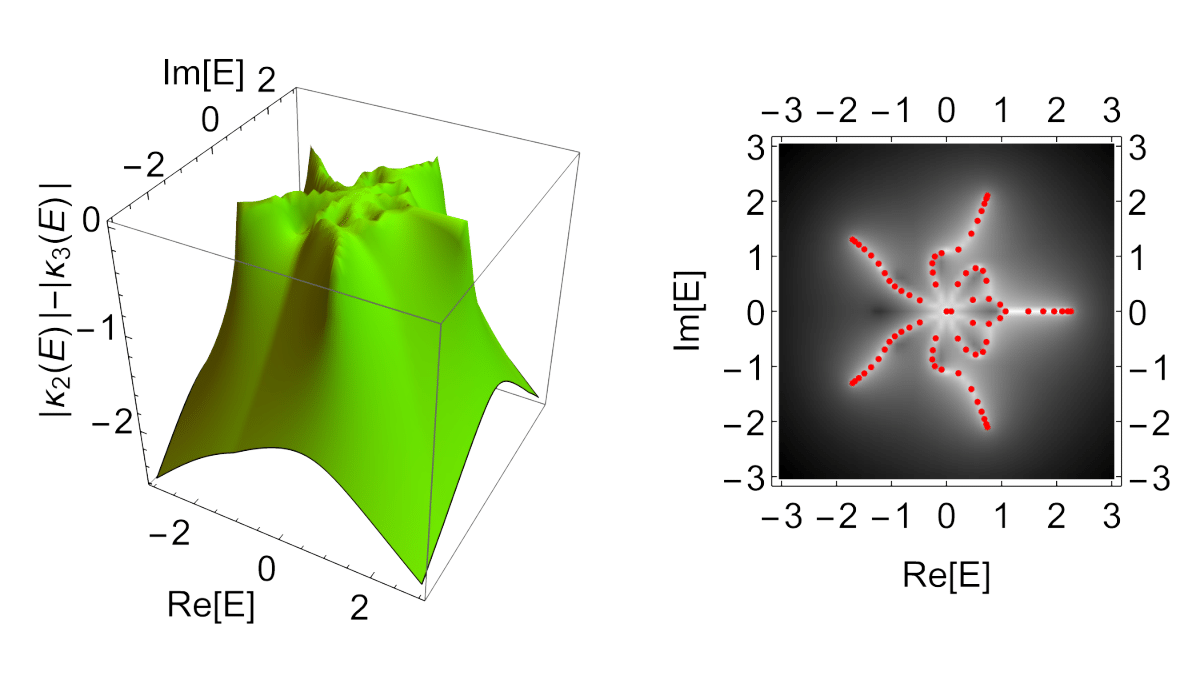}}
\end{minipage}

\caption{\textbf{Inverse skin depth surface intersections illustrate OBC spectra.} Under open boundary conditions (OBC), the eigenenergies $\overline{E}$ (red dots) occur at the inverse skin depth surface intersections $|\kappa_i(E)|$ and $|\kappa_j(E)|$ (the white trajectories traced out in the complex energy plane). $\kappa$ defines the Generalized Brillouin Zones (GBZ) of the system, where $\kappa=\text{Im}(k)$. The bands $i,j$ are chosen to give the smallest $|z|$ solution. This is illustrated for systems with energy dispersions given by the characteristic polynomials $P(z,E)=0$. Two examples shown here are (a) simple 3-fold symmetric OBC spectrum: $P(z,E)=z^2+1/z-E$ and (b) more sophisticated OBC spectrum: $P(z,E)=z^2+1/z^2-2E^2z-E^3$. Since the OBC eigenenergy computations obtained via diagonalization quickly become slow and numerically inaccurate, we will from now on mostly use these surface intersection plots as surrogates for the OBC spectra. }
\label{fig:S1}
\end{figure}

Note that in Hermitian and reciprocal non-Hermitian lattices, there already exists two solutions to Eq.~\ref{CP} for $\text{Im}(k)=0$. For the former, this is self-evident from the fact that $E(k)$ generically visits every $E$ point at least twice as $k$ varies periodically from $0$ to $2\pi$. Hence the OBC and PBC spectra of large Hermitian and reciprocal non-Hermitian lattices approximately agree, with topological mid-gap states being isolated exceptions. However, in generic lattices with arbitrary hopping coefficients in either directions, Eq.~\ref{CP} is likely satisfied only by two energy-degenerate solutions of equal $\text{Im}(k)\neq 0$. Hence the PBC and OBC spectra are totally different, a phenomenon known as the non-Hermitian skin effect (NHSE). Under OBC, eigenstates due to NHSE can exponentially localize at a boundary. Below, we shall frequently employ the inverse skin depth surface intersections (demarcated in white) to illustrate the OBC spectra (red dots). Two comparison examples are given in Fig.~\ref{fig:S1}.

\section{Separable cases}

\begin{figure}
\centering
\includegraphics[width=\linewidth]{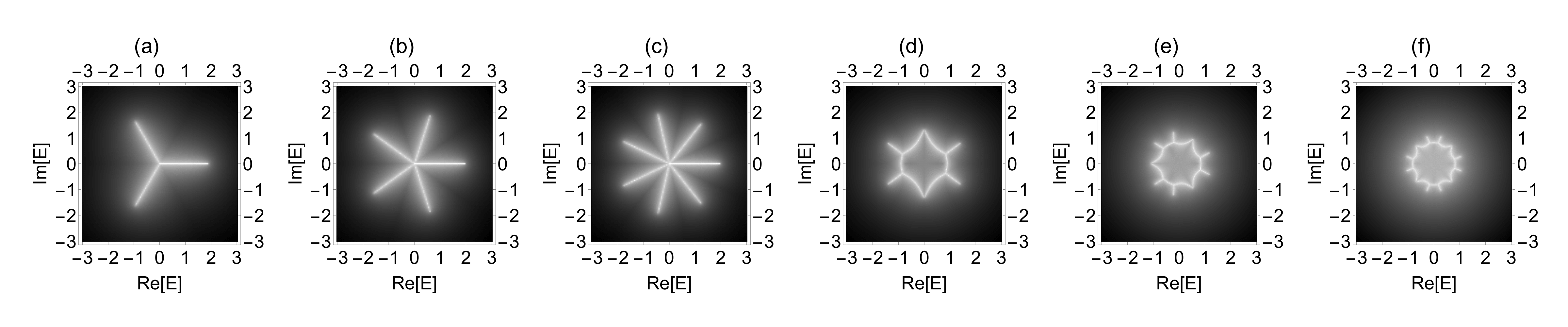}
\caption{\textbf{Examples of OBC spectra.} \textbf{(a-c)} For a separable characteristic polynomial (Eqn.~\ref{sep}) with $f=1$, the OBC spectrum $\bar E$ takes the shape of a star with $\mathcal{N}=(a + b)$ number of branches, where (a) $\mathcal{N}=3$, (b) 5 and (c) 7. \textbf{(d-f)} Multiple copies of the (distorted) OBC spectral loci can be obtained through conformal mappings in the complex energy plane. They form closed loops when adjacent copies intersect. Depicted are the OBC spectra $\overline{E}$ consisting of (d) $N=2$, (e) $3$ and (f) $4$ copies of the $N=1$ case, corresponding to $N$-band Hamiltonians of Eq.~\ref{multiHz} which are characterized by the characteristic polynomial $P(E,z)=(z^3 - 2z^{-1} + 2z^{-2})-E^N$.}
\label{fig:S2}
\end{figure}

We first consider the most straightforward case where $P(E,z)=0$ can be written such that $E$ and $z$ appears separately on either side of the equation. We write that as $P(E,z)=F(E)-p(z)=0$ i.e. 
\begin{equation}
F(E)=p(z).
\end{equation}
Algebraically, this separable case is rather easy to handle, because for every given $E$, the form of the polynomial $p(z)$ stays the same. First to be examined is the simplest nontrivial case where $p(z)$ consists of two terms:
\begin{equation}\label{sep}
F(E)=c_1z^a+\frac{c_2}{z^b} 
\end{equation}
where $a,b>0$, $z=e^{ik}$ and $f=\deg(F)$ is the degree of the polynomial $F(E)$. Physically, this polynomial corresponds to the 
1-band non-reciprocal Hamiltonian with forward hoppings $c_1$ across $a$ sites, and reverse hoppings $c_2$ by $b$ number of sites. For this case, it has been shown that~\cite{lee2020unraveling} the OBC spectrum $\bar E$ takes the shape of a star with $\mathcal{N}=(a + b)f$ number of branches, as shown in Fig.~\ref{fig:S2}(a-c). This result can also be intuitively obtained by regarding $a+b$ as the number of times the Brillouin zone (BZ) is folded. 

Notably, the exact values of the hopping coefficients $c_1$, $c_2$ do not affect the graph topology of the OBC spectrum, since they merely rescale the asymmetry of the hoppings. To see this, we rescale $z=e^{ik}\rightarrow\alpha w$, such that
\begin{equation}
F(E)=c_1\alpha^a\bigg[w^a+\frac{c_2}{c_1}\alpha^{-(a+b)}\frac{1}{w^b}\bigg] 
\end{equation}
where we have chose $\alpha=(c_1/c_2)^{a+b}$, with the multiplicative factor $c_1\alpha^a$ inconsequential. Depending on our choice of $\alpha$, we can obtain any relative hopping coefficient with the square parentheses, even though the graph topology of the OBC spectrum remains unchanged. 

From this well-understood case, one can extend our understanding to an entire class of other cases that are related via a conformal transformation of $E$, i.e. $F(E)\rightarrow F'(E)$. 
Specifically, via the mapping $E\rightarrow E^N+d$, we can map an arbitrary dispersion $E=p(z)$ of a single-component lattice into the dispersion $E^N+d=p(z)$ of a $N$ by $N$ Hamiltonian of the form
\begin{equation}
H(z)=\begin{pmatrix}0&p(z)-d\\\text{Id}_{N-1}&0\\\end{pmatrix}.
\label{multiHz}
\end{equation}
The resulting OBC spectrum will consist of $N$ number of replicas of the corresponding OBC spectrum of the single-component case in complex $E$-space, unique up to the conformal deformation, perturbed by the translation $d$. As a cautionary note, this conformal mapping argument can only be applied to $E$ and not $z$, since the powers of $z$ describe hopping distances which may be truncated under OBCs.

With clever choices of conformal mappings, one may create OBC spectra with interesting polygonal shapes and closed loops. We demonstrate some examples in Fig.~\ref{fig:S2}(d-f), where we consider $E=p(z)=z^3+2/z^2-2/z\mapsto E^N=p(z)$ for $N=2,3,4$. The single component spectrum gets mirrored into $N$ copies and since each of these copies terminate at lines of symmetry (regularly spaced by $2\pi/N$ in the complex $E$-plane), the OBC spectrum $\overline{E}$ of the resultant $N$-component Hamiltonian traces a closed loop with $N$-fold symmetry in the complex $E$ plane.

Naturally, one will ask how would the OBC spectrum change, in the presence of multiple asymmetric hopping terms such that in one direction, we already have two or more competing hopping terms. For concreteness, we add the term $z^c$ in the characteristic polynomial, i.e. 
\begin{equation}\label{poly}
P(E,z)=z^a+1/z^b+r\,z^c-F(E) 
\end{equation}
where there is no restriction on the sign of $c$, although we impose $|c|<a,b$ without loss of generality (if $c$ is greater, we can always trivially rearrange the polynomial terms). To see the effect of this extra term, we modulate the strength of the perturbation $r$, keeping $r$ real. Again, the coefficient of either $z^a$ or $1/z^b$ merely rescales the system and thus does not matter.  Consider the following example 
\begin{equation}\label{polyspecial}
P(E,z)=z^2+1/z+r\,z-E. 
\end{equation}
If $r=0$, we recover the previous result and the OBC spectra $\overline{E}$ lie along a star with $\mathcal{N}=(2+1)\times 1=3$ branches. If $r\neq 0$ but small, the branches will be expected to distort slightly. The exact analytical derivation of this distortion is difficult considering that one have to solve a characteristic polynomial cubic in $z$, followed by imposing the GBZ condition of degenerate $|z|$ and $E$, which results in a complicated mess of terms. But for large $r$, we can use an asymptotic argument to deduce the qualitative appearance of the OBC spectrum. In this limit, Eq.~\ref{polyspecial} reduces to
$P(E,z)\sim r~z+1/z-E$, which is none other than that of the well-known Hatano-Nelson model~\cite{hatano1998non,hatano1996localization}. In what follows, we will show that the OBC spectrum $\overline{E}$ in this asymptotic limit is purely real or purely imaginary (effectively 2 wedge star), depending on the sign of $r$. We proceed to solve for the OBC eigenenergies in this asymptotic limit:
\begin{equation}
0=rz+1/z-E\implies z=\frac{-E\pm\sqrt{E^2-4r}}{2r}
\end{equation}
To obtain the OBC eigenvalues $\overline{E}$, we impose the GBZ condition~\cite{lee2019anatomy} to obtain $(\bar E^2-4r)=-s\bar E^2$ for some $s\in\mathbb{R}$ which gives  $\bar E^2=\frac{4r}{1+s}\in\mathbb{R}$. Hence, for large $r$, $\bar E$ can either be purely real or purely imaginary and this depends on $\sgn(r)$, i.e. if $r<0$, $i\bar E\in\mathbb{R}$. For intermediate values of $r$, we can deduce the form of the OBC spectra from our knowledge of the large $r$ limit and $r=0$ limit. To obtain the exact OBC modes, we either compute this numerically or solve $P(E,z)$ which is cubic in $z$. 

\begin{figure*}
\centering
\includegraphics[width=\linewidth]{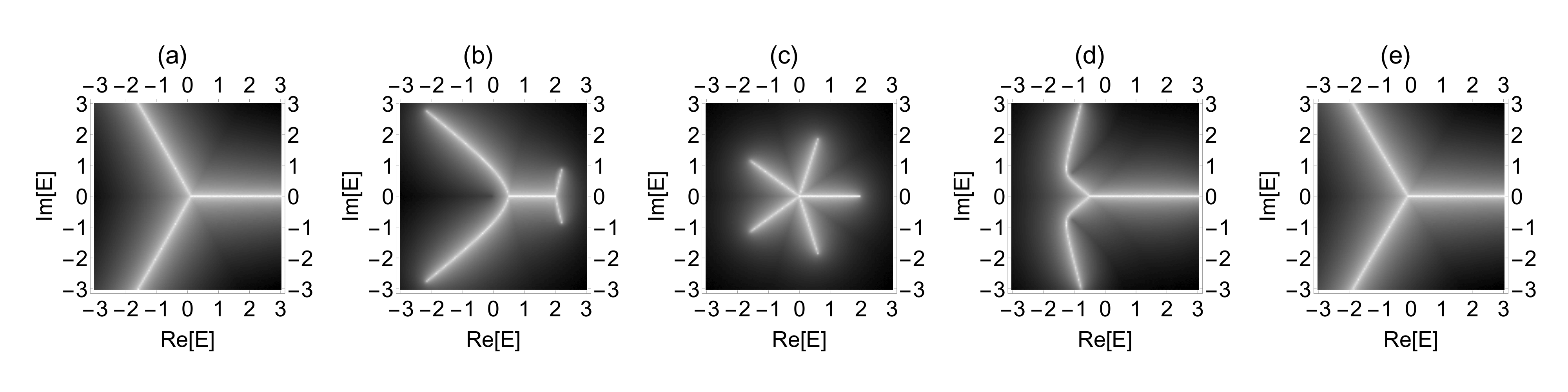}
\caption{\textbf{Competition between hoppings and spectral interpolation between limiting cases.} OBC spectra $\overline{E}$ corresponding to a Hamiltonian with characteristic polynomial $P(E,z)=z^3+1/z^2+ rz -E$. $r$ describes the competition between two hoppings along a certain direction, thus the OBC spectra interpolates between a 3-star and a 5-star as $r$ sweeps from: (a) $r=-10$, (b) $r=-2$, (c) $r=0$, (d) $r=2$, (e) $r=10$.}
\label{fig:S3}
\end{figure*}

For a slightly more sophisticated example with competing hoppings along a certain direction, consider: 
\begin{equation}\label{polyspecial2}
P(E,z)=z^3+1/z^2+r\,z-E 
\end{equation}
At $r=0$, we recover a 5-star from $\mathcal{N}=3+2=5$. At large $|r|$, Eq.~\ref{polyspecial2} becomes $P(E,z)\sim r~z+1/z^2-E$, which upon solving for the OBC states, we recover a 3-star as expected, see Fig.~\ref{fig:S3}. At intermediate values of $r\neq 0$, some of the branches shrink and grow accordingly such that we can interpolate between the small and large $|r|$ cases.
 
In general, when there is more than one competing hoppings in a certain direction, we can neglect all the smaller hoppings if one hopping coefficient is much larger than the rest. By interpolating between such limits, we can deduce the qualitative form of the OBC spectrum $\overline{E}$ without the explicit construction of the generalized Brillouin zone.

\newpage
\section{Non-separable cases}

More generally, the bivariate characteristic polynomial contains mixed terms
\begin{equation}\label{bivar}
P(E,z)=\sum_{i,j}c_{i,j}E^iz^j,
\end{equation}
and cannot be separated into a sum of univariate polynomials of $E$ and $z$. Indeed, in the presence of mixed terms like $E^pz^q$ for some $p,q\neq 0$, the OBC spectra $\overline{E}$ is significantly harder to be solved analytically. However, as we will show, we may use our previous asymptotic arguments to help deduce the OBC spectral branching patterns of these non-separable cases under certain regimes.

\subsection{One hopping length}
As a warmup, we consider a bi-variate polynomial Eq.~\ref{bivar} with only one hopping length (unique order of $z$ or $z^{-1}$), but in the presence of mixed terms $E^iz^j$ for some $i,j\neq 0$: 
\begin{equation}\label{onelength}
P(E,z)=\frac{1}{z}+rG(E)z-F(E) 
\end{equation}
where $r\in\mathbb{R}$ controls the strength of the perturbation induced by the mixed term, and $G(E)$, $F(E)$ are arbitrary functions of $E$. As Eq.~\ref{onelength} is effectively quadratic in $z$, this is analytically tractable when we impose the GBZ condition. However, we will show that its OBC spectrum $\overline{E}$ is not too exciting, and independent of $r$. That said, its solution will be invaluable to the analysis of other more complicated examples that follow. Characteristic polynomials of the form of Eq.~\ref{onelength} can be obtained via suitably designed multi-component Hamiltonians: consider for instance the simplest case of $F(E)=E^2$ and $G(E)=E$, which corresponds to the 2-band non-reciprocal Hamiltonian
\begin{equation}
H_{\substack{F(E)=E^2,\\ G(E)=E}}(z)=\begin{pmatrix} rz&1/z\\1&0\\\end{pmatrix}.
\end{equation}

For arbitrary $F(E)$ and $G(E)$ and $r\neq 0$, we can solve for $z$ and apply the GBZ condition of having the $\pm$ solutions being of equal $|z|$ magnitudes:
\begin{align}
z&=\frac{F(E)\pm\sqrt{F(E)^2-2rG(E)}}{2rG(E)}\nonumber\\&=\frac{1}{2r}\bigg(\frac{F(E)}{G(E)}\pm\sqrt{\frac{F(E)^2}{G(E)^2}-\frac{2r}{G(E)}}\bigg)\nonumber\\\implies &\bigg(\frac{F(E)}{G(E)}\bigg)^2-\frac{2r}{G(E)}=-s^2\bigg(\frac{F(E)}{G(E)}\bigg)^2,~s\in\mathbb{R} 
\end{align}
Further simplifications give $F(E)^2/G(E)=\frac{4r}{1+s^2}\in\mathbb{R}$, consistent with Eqn.~\ref{real}. Keeping $F(E)$ and $G(E)$ to be strictly monomials, the OBC spectra will be of star-shape form, with number of branches being
\begin{equation}
\mathcal{N}=2f-g 
\end{equation}
where $f=\deg(F(E))$ and $g=\deg(G(E))$. Note that the Hamiltonian that gives rise to the requisite characteristic polynomial need not be unique. Consider a more sophisticated example - a three-band Hamiltonian, i.e. with $F(E)=E^3$ and possible choices for $G(E)$ being $E^2$ and $E$ respectively:
\begin{equation}
H_{\substack{F(E)=E^3,\\ G(E)=E^2}}(z)= \begin{pmatrix}0&rz&1/z\\1&0&0\\0&1&0\\\end{pmatrix},\quad H_{\substack{F(E)=E^3,\\ G(E)=E}}(z)=\begin{pmatrix}rz&0&1/z\\1&0&0\\0&1&0\\\end{pmatrix} 
 \end{equation}
From the above results, the OBC eigenenergies satisfy $\overline{E}^5=\frac{2r}{1+s^2}\in\mathbb{R}$ and $\overline{E}^4=\frac{2r}{1+s^2}\in\mathbb{R}$ respectively, hence giving rise to a five-star and four-star respectively.

\begin{figure}
\centering
\includegraphics[width=\linewidth]{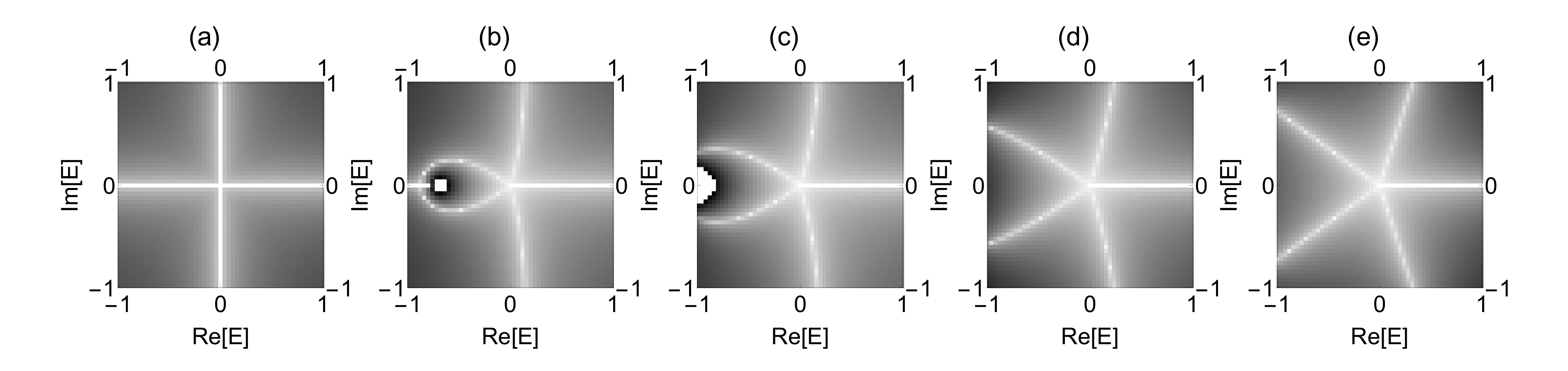}
\caption{
\textbf{Interpolation between dominant hopping terms. }
For the characteristic polynomial $P(E,z)=1/z+(r_1 E^2+r_2E)z-E^3$, with (a) $r_1=1$, $r_2=0$, (b) $r_1=1$, $r_2=0.7$, (c) $r_1=1$, $r_2=1$, (d) $r_1=0.5$, $r_2=1$, (e) $r_1=0$, $r_2=1$; one can interpolate between a $E^2z$-dominant OBC spectrum (4-star in (a)) and a $E z$-dominant spectra (5-star in (e)) by tuning the relative ratio of $r_1$ and $r_2$. }
\label{fig:S4}
\end{figure}

To justify that we do not miss out qualitatively new behavior by restricting $F(E)$ and $G(E)$ to monomials, we consider the previous three-band Hamiltonian example, but with a linear superposition of the two possible choices for $G(E)$: $G(E)=r_1E^2+r_2E$. The corresponding Hamiltonian is
\begin{equation}
    H_{\substack{F(E)=E^3,\\ G(E)=r_1E^2+r_2E}}(z)=\begin{pmatrix}r_2z&r_1z&1/z\\1&0&0\\0&1&0\\\end{pmatrix} 
\end{equation}
The limiting cases give back 5-star and 4-star respectively for $r_1=0$, $r_2=1$ and $r_1=1,$ $r_2=0$. Interpolating between the two cases, the branches deform accordingly. As $r_1/r_2$ increases from 0 to 1, two of the five branches in $\text{Re}[E]<0$ bend towards each other and fuse into one branch. As $r_1/r_2>1$, we have these two branches combine such that the total branches becomes 4 as $r_2\rightarrow 0$. Simultaneously, the two near-vertical branches become purely imaginary. See Fig.~\ref{fig:S4}. Hence, for the remaining of this work, to understand the effect solely due to the mixed terms $E^iz^j$, it is thus reasonable for us to strictly consider monomials in $E$. 

\subsection{Two hopping lengths - Canonical Characteristic Polynomial}
After the preceding warm-up examples, we now discuss the crux of this work, which concerns a broad class of characteristic polynomials of the following canonical form which is a great generalization of Eq.~\ref{onelength}: 
\begin{equation}\label{polycrux}
P(E,z)=Q(z)+r~G(E)J(z)-F(E). 
\end{equation}
This is the general ansatz for $P(E,z)$ consisting of the sum of a $z$-dependent term, a $E$-dependent term, and a cross term that is factorizable in terms of a $E$ and a $z$-dependent term. Due to the determinant definition of the characteristic polynomial, $\text{deg}(F) > \text{deg}(G)$ must be true.

As we have shown previously with different examples, consideration of monomials for $J(z)$ is sufficient. We first consider $\deg(Q)>\deg(J)$ and 
$Q(z) =z^a+1/z^b$, representing two distinct hopping lengths. If $Q(z)$ contains additional competing hopping terms, then from previous examples, we will study them by interpolating between extreme cases where only two terms dominate, one in each direction. A possible choice of a Hamiltonian with this characteristic polynomial given by Eq.~\ref{polycrux} is
\begin{equation}
H(z)=\begin{pmatrix}A&Q(z)\\\text{Id}_{f-1}&0\\\end{pmatrix},
\end{equation}
where $f=\deg(F)$, and $A$ is a row matrix with $f-1$ entries, such that the $g-$th entry of $A$ is $rJ(z)$ where $g=\deg(G)$. By construction, $g<f$. In general, analytically solving the OBC eigenenergies $\overline{E}$ is difficult, if not impossible for very high orders of $Q(z)$, $H(z)$. However, under certain limits, the asymptotic argument used earlier will be helpful again. For monomials $F(E)$, $G(E)$, the main possibilities are as follows:

\begin{enumerate}
    \item $r=0$: the polynomial Eq.~\ref{polycrux} reduces to the familiar form $P(E,z)=Q(z)-F(E)$ of the same type as Eq.~\ref{poly} and as previously shown, the OBC spectrum $\overline{E}$ is a star with
    \begin{equation}\label{N0}
   \mathcal{N}_0=(a+b)f
   \end{equation}
	number of branches, where $a$ and $b$ are the orders of the dominant left and right hoppings in $Q(z)$. Specifically, the OBC eigenenergies $\overline{E}$ lie along lines with constant angular spacings of $\theta=\frac{2\pi}{\mathcal{N}_0}$ where $E=E_0e^{i\theta}$.
	
    \item $r\neq 0$, for sufficiently small $|E|$: since $g<f$, $F(E)$ decays more rapidly, so Eq.~\ref{polycrux} has the asymptotic small $|E|$ behaviour of $P(E,z)\sim Q(z)+rG(E)J(z)$. Interestingly, the OBC spectrum $\overline{E}$ bears the familiar branching form with the number of branches being
    \begin{equation}\label{NS}
    \mathcal{N}_S=(a+b)g\quad\forall r\neq0
    \end{equation}
    Again, the OBC eigenenergies $\overline{E}$ lie along lines with constant angular intervals of $\frac{2\pi}{\mathcal{N}_S}$. The precise form of monomial $J(z)$ does not matter for this case, since any power of $z$ can be absorbed by the $Q(z) =z^a+1/z^b$ term.
    
		\item For sufficiently large $r\neq 0$ as well as large $|E|$: we need to compare the asymptotic behaviours of  $Q(z)$ and $J(z)$ together. To determine which term in $Q(z)=z^a+1/z^b$ dominates, we inspect the sign of $j=\deg(J)$. Concretely, if $j>0$, then asymptotically, Eq.~\ref{polycrux} tends to $P(E,z)\sim 1/z^b+rG(E)J(z)-F(E)$. Similarly, if $j<0$, then asymptotically $P(E,z)\sim z^a+rG(E)J(z)-F(E)$. To determine the number of branches, suppose the dominating term in $Q(z)$ is $z^q$ for some $q\neq0$ and suppose further $j<0$, then by keeping only the dominating $z$ and $E$ terms, the characteristic polynomial becomes
		\begin{equation}
		P(E,z)\sim z^{|q|} + r E^g z^{-|j|} - E^f.
		\label{PEzfgh}
		\end{equation}
	To find the number of branches $\mathcal{N}_L$, suppose that $(E,z)$ is a solution to $P(E,z)=0$. Other solution branches are related to it via simultaneous phase rotations $E\rightarrow Ee^{i\theta}$, $z\rightarrow ze^{i\phi}$. Subtituting that into Eq.~\ref{PEzfgh}, we have 
\begin{equation}
e^{i|q|\phi}\left(z^{|q|} +r E^gz^{-|j|}e^{i(g\theta-(|j|+|q|)\phi)}\right)=e^{if\theta}E^f
\end{equation}
For this to be consistent with $P(E,z)=0$ i.e. $z^{|q|} + r E^g z^{-|j|} = E^f$, we need
 $g\theta = (|j|+|q|)\phi+2n\pi$ and $f\theta - |q|\phi = 2m\pi$, where $n,m$ are integers. Solving, we obtain
\begin{equation}
\theta = \frac{2\pi n'}{-g|q|+(|q|+|j|)f}
\end{equation}
where $n'=-n|q|+m(|j|+|q|)=\text{GCD}(|q|+|j|,|q|)$ is the greatest common divisor of $|q|+|j|$ and $|q|$. The case with $j>0$ shall proceed analogously with the same result. Hence the number $\mathcal{N}_L=\frac{2\pi}{\theta}$ of different branches of $E$ that are related to the original branch via $(E,z)\rightarrow (Ee^{i\theta},ze^{i\phi})$ is given by
\begin{equation}\label{NL}
    \mathcal{N}_L=\frac{(|q|+|j|)f-|q|g}{\text{GCD}(|q|+|j|,|q|)}.
\end{equation}
Approximately, the OBC eigenenergies $\overline{E}$ lie along lines with constant angular intervals of $\theta = \frac{2\pi}{\mathcal{N}_L}$.
\end{enumerate}
These rules will be exemplified by several detailed examples later. For other parameter ranges of $r$ and $|E|$, one may deduce the OBC spectrum $\overline{E}$ by interpolating from the results obtained from the above limiting cases.

\newpage
\subsection{Global symmetry properties of OBC spectra}

Here we discuss some general symmetry properties of the OBC spectrum of our canonical characteristic polynomial Eq.~\ref{polycrux}. For convenience, we restate its expression again:
\begin{equation}
P(E,z)=Q(z)+r~G(E)J(z)-F(E)\label{globalsym}
\end{equation}
with $Q(z)=z^a+\frac1{z^b}$, $f=\text{deg}(F)$, $g=\text{deg}(G)$ and $j=\text{deg}(J)$. As argued before, it is sufficient to consider $F(E)$ and $G(E)$ to be strictly  monomials in $E$, and that $Q(z)$ has the form $z^a+1/z^b$. We have seen previously that the number of branches in the various regimes are approximately separated by constant angular intervals where we find this number via simultaneous phase rotations $E\rightarrow Ee^{i\theta}$, and $z\rightarrow ze^{i\phi}$. \\

\begin{figure}
    \centering
    \includegraphics[width=0.8\linewidth]{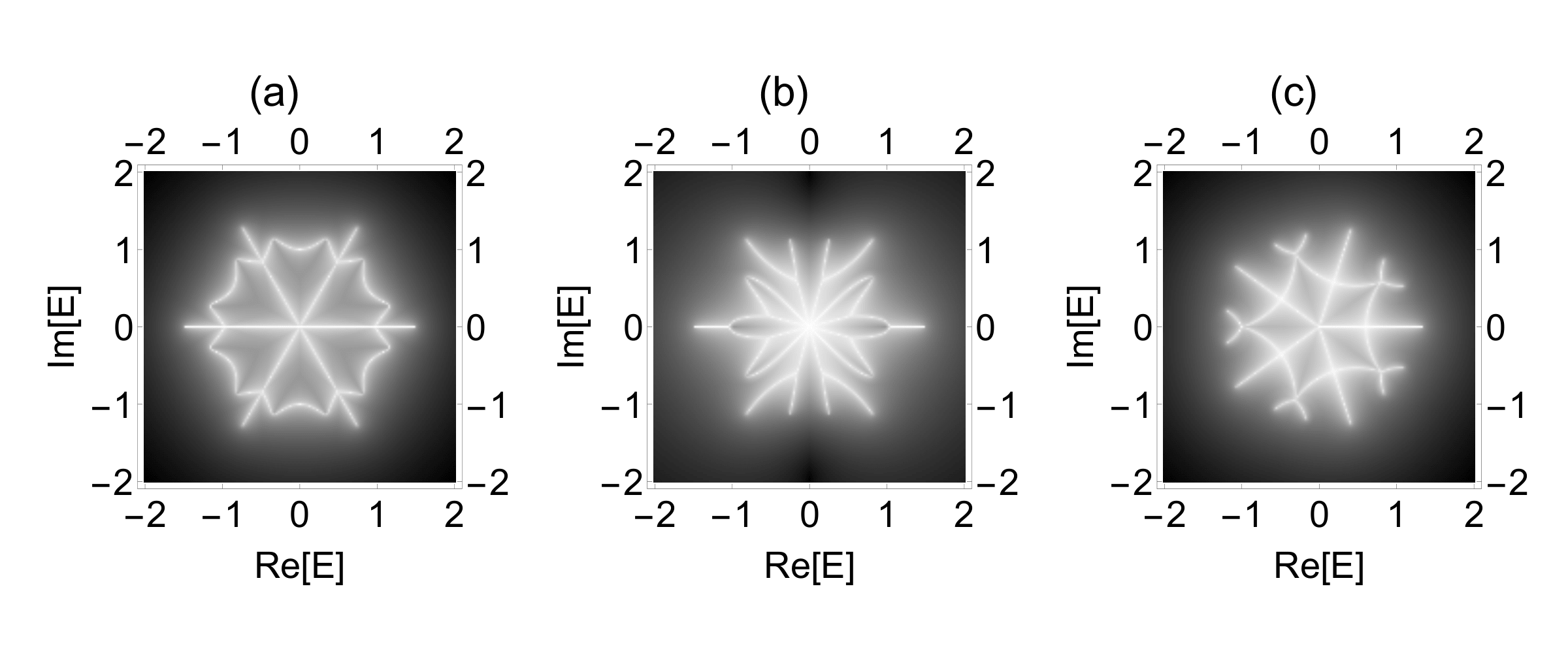}
    \caption{\textbf{Examples of OBC spectra with global symmetry.} Eqn~\ref{globalsym} has $r=1$ (a-c) and $r=-1$ (d). \textbf{(a)} 6-fold symmetry for $Q(z)=z^3+z^{-3}$, $J(z)=z^2$, $G(E)=E$, $F(E)=E^3$; \textbf{(b)} reflection symmetry for $Q(z)=z^2+z^{-2}$, $J(z)=z$, $G(E)=E^3$, $F(E)=E^4$; \textbf{(c-d)} 5-fold symmetry for $Q(z)=z^3+z^{-2}$, $J(z)=z$, $G(E)=E$, $F(E)=E^4$.}
    \label{fig:S5}
\end{figure}

(1) When the spectrum has global $n$-fold rotation, we can find a relation between $\theta$ and $\phi$. Specifically, $\theta$ will be $2\pi/n$ and hence $\phi$ is a function of $n$ and it is manifested as a permutation of the OBC modes. If the characteristic polynomial $P(E,z)$ is invariant to this simultaneous phase rotations $\forall\overline{E}$, i.e.
\begin{equation}
P(\overline{E},z)\rightarrow P(\overline{E}e^{i2\pi/n},ze^{i\phi(n)})=P(\overline{E},z)\label{n-fold}
\end{equation}
then the OBC spectrum $\overline{E}$ is said to have $n$-fold global symmetry. Since the rotational symmetry extends $\forall\overline{E}$, both leading terms in $Q(z)$ are equally important and we require $n$ to be equal to $a+b$. To explicitly work out the conditions for $n$-fold global symmetry, we solve Eqn.~\ref{n-fold}. The form of Eqn.~\ref{globalsym} will yield three inter-related conditions
\begin{equation}
 e^{i(a+b)\phi}=1,\quad e^{i(b+j)\phi}e^{ig2\pi/n}=1,\quad e^{i(f2\pi/n+b\phi)}=1
\end{equation}
The existence of a unique solution for $\phi$ will guarantee the OBC spectra to have $n$-fold symmetry. Moreover, suppose $\gcd(f,g)=m>1$, then we must have $\phi=2\pi m/n$. Equivalently, this is achieved by conformally transforming both $F(E)$ and $G(E)$ via $E\rightarrow E^m$. To illustrate this concretely, consider the example of 
\begin{equation}
P(E,z)=z^3+\frac{1}{z^3}+r z^2G(E)-F(E)
\end{equation}
Suppose we have $G(E)=E$ and $F(E)=E^3$, it follows that $\theta=\phi=\pi/3$ and thus the spectrum has 6-fold symmetry, as illustrated in Fig.~\ref{fig:S5}a. By performing  a conformal transform $E\rightarrow E^3$, one can obtain $G(E)=E^3$ and $F(E)=E^9$, in turn we have a spectrum with 18-fold symmetry, hence we recover $\theta=\pi/9$ and $\phi=\pi/3$. We can generalize this to higher powers of $f$ and $g$.\\

A special case would be inversion symmetry about the imaginary $E$ axis, which is equivalent to 2-fold rotational symmetry ($\theta=\pi$). One such example is
\begin{equation}
P(E,z)=z^2+1/z^2+r E^3z-E^4
\end{equation}
where the corresponding OBC spectra $\overline{E}$ has reflectional symmetry about the $\text{Im}(E)$ axis (Fig.~\ref{fig:S5}b).\\

(2) The spectrum can also have inversion symmetry about the imaginary $E$ axis under inversion of $\sgn(r)$. This is what we call $r$-symmetry. The characteristic polynomial will have to be invariant to the negation of $r$:
\begin{equation}
P(\overline{E},z;r)\rightarrow P(\overline{E}e^{i\theta},ze^{i\phi};-r)=P(\overline{E},z;r)\label{r-sym}
\end{equation}
where $\overline{E}e^{i\theta}$ has the real component of $\overline{E}$ negated. For any complex number $c=|c|e^{i\chi}\in\mathbb{C}$, negating its real part is equivalent to a gauge transformation:
\begin{equation}
-c^*=-|c|e^{-i\chi}=-ce^{-i2\chi} = ce^{i(\pi-2\chi)}\label{negation}
\end{equation}
for some phase $\chi$ that satisfies $c=\cos\chi+i\sin\chi$ by Euler's formula. Hence, Eqn.~\ref{negation} gives $\theta = \pi-2\chi$. Again, explicitly solving Eqn.~\ref{r-sym} will give three inter-related conditions
\begin{equation}
    e^{i(a+b)\phi}=1,\quad e^{i(b+j)\phi}e^{i(-2\chi g+(g+1)\pi)}=1,\quad e^{if(\pi-2\chi)}e^{ib\phi}=1
\end{equation}
The existence of a unique solution for $\phi$ and $\chi$ will guarantee this symmetry. Sometimes, this mirror symmetry under the inversion of $r$ may also appear together with the $n$-fold rotation symmetry. One such example is
\begin{equation}
P(E,z)=z^3+\frac{1}{z^2}+rE z-E^4
\end{equation}
where its OBC spectrum $\overline{E}$ is also invariant under a $\pi/5$ rotation (Fig.~\ref{fig:S5}c).

\section{Detailed analysis of illustrative examples with $Q(z)=z^2+1/z$ and $J(z)=z$}

Here we provide in-depth analysis to some non-trivial illustrative examples.

To have a more concrete understanding, we consider the simplest non-trivial form  $Q(z)=z^2+1/z$ and $J(z)=z$.
\begin{equation}\label{polysimp}
P(E,z)=z^2+\frac{1}{z}+rG(E)z-F(E)
\end{equation}
This, however, is still too difficult for determining the OBC modes analytically. While we may be able to demonstrate the 3-star when $r=0$ via Cardano's formula, it is generally difficult for arbitrary $r$. For monomials $G(E)=E^g$ and $F(E)=E^f$, we have three distinct cases, namely
\begin{enumerate}
    \item $r=0$: Eq.~\ref{N0} gives the number of branches to be $\mathcal{N}_0=3f$ for all OBC eigenenergies $\overline{E}$;
    \item $r\neq 0$, small $|\overline{E}|$: the characteristic polynomial Eq.~\ref{polysimp} behaves asymptotically like $P(E,z)\sim z^2+1/z+r G(E)z$. From Eq.~\ref{NS}, we then have the number of branches to be $\mathcal{N}_S=3g$;
    \item for sufficiently large $r\neq 0$, large $|\overline{E}|$: the characteristic polynomial Eq.~\ref{polysimp} behaves asymptotically like $P(E,z)\sim 1/z+r G(E)z-F(E)$, which reduces to the previous analytic example. Hence, the number of branches is $\mathcal{N}_L=2f-g$, consistent with that obtained in Eq.~\ref{NL}.
\end{enumerate}
Next, we illustrate the above three cases by demonstrating different examples of $G(E)$ and $F(E)$:

\begin{figure*}
\centering
\includegraphics[width=\linewidth]{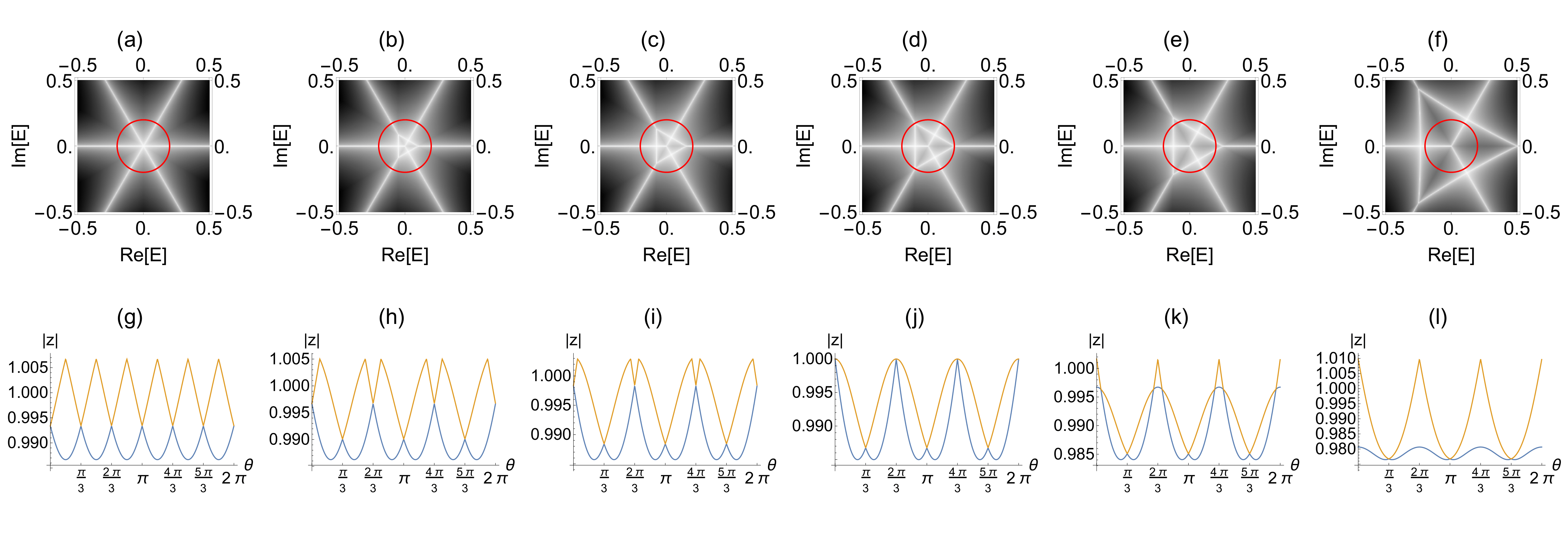} 
\caption{\textbf{Example 1.} For the polynomial $P(E,z)=z^2+1/z+r E z -E^2$, with  $r=0,-0.1,-0.15,-0.2,-0.25$ and $-0.5$ for (a-f) and (g-l). 
\textbf{(a-f)} Evolution of the OBC triangular spectra obtained only from a finite small perturbation, $0<|r|<1$. \textbf{(g-l)} We probe the growing triangle with a constant circle of radius $E_0=0.2$. \textbf{(g)} When the triangle is not captured, all $\mathcal{N}_0=6$ branches are manifested as inverse skin depth surface intersections, i.e. $n=6$. \textbf{(h-k)} When the triangle is slightly bigger, the circle intersects at two points in the vicinity of each triangle vertex, giving us a total of $n=9$, inclusive of the 3 radial branches. \textbf{(l)} On the other hand, when the triangle is much bigger, the circle captures only the 3 radial branches, i.e. $n=\mathcal{N}_S=3$.}
\label{fig:S6}
\end{figure*}

\subsubsection{Example 1: $G(E)=E$, $F(E)=E^2$}

Fig.~\ref{fig:S6}(a-f) shows the OBC spectra $\overline{E}$ for $G(E)=E$ and $F(E)=E^2$ for various $r$ values. When $r=0$, we have two copies of the single component spectrum, giving a 6-star, as expected (Eqn.~\ref{N0}). The inverse skin depth surfaces $|z_1(E)|$ and $|z_2(E)|$ appear periodic with period $\frac{2\pi}{6}=\pi/3$. When $r\neq0$ is small, one of the copy transforms to an equilateral triangle centred at the origin. This ensures the number of radial branches for small $|E|$ is kept at $\mathcal{N}_S=3$, as predicted by Eqn.~\ref{NS}. The periodicity of the inverse decay length doubles to $2\pi/3$, with the `mirror' star from the 2-component system to exhibit a deeper well, i.e. smaller $|z|$.  The periodicity of the inverse decay length and the spectrum is consistent with the periodicity of the characteristic polynomial $P(E,z)$. 

\begin{figure*}
\begin{minipage}{\linewidth}
\subfloat{\includegraphics[width=\linewidth]{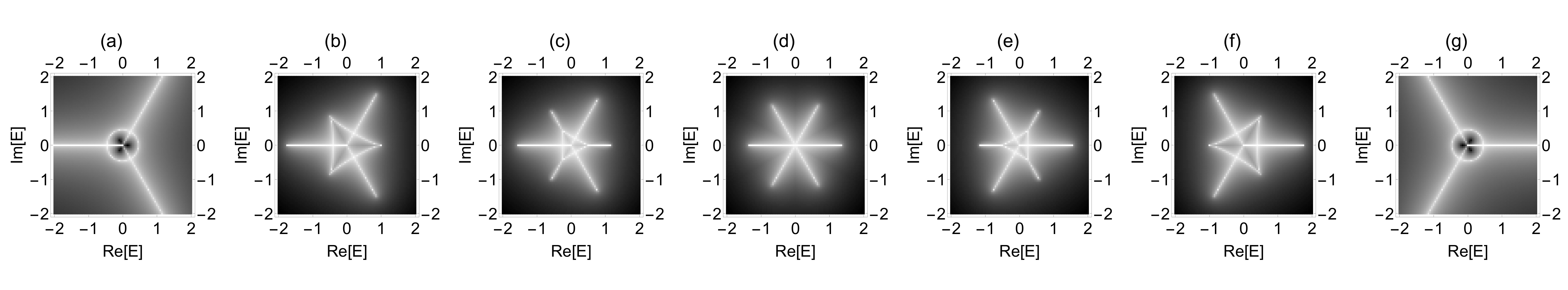}}\\
\subfloat{\includegraphics[width=\linewidth]{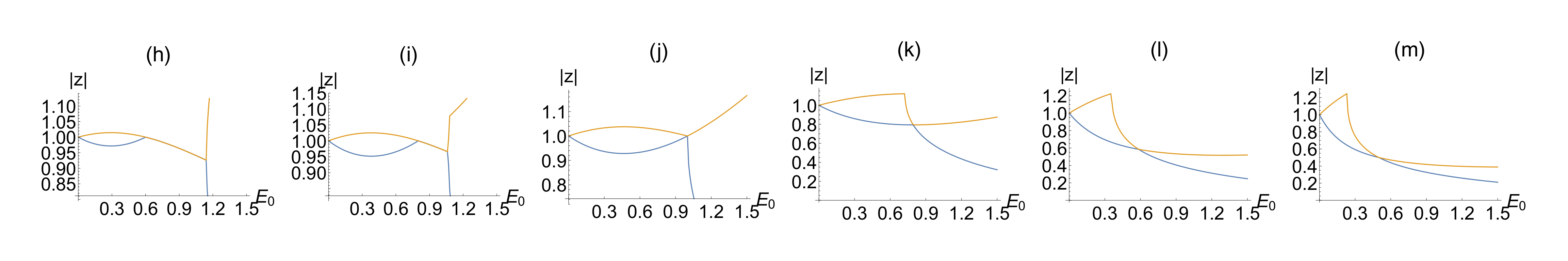}}\\
\subfloat{\includegraphics[width=0.6\linewidth]{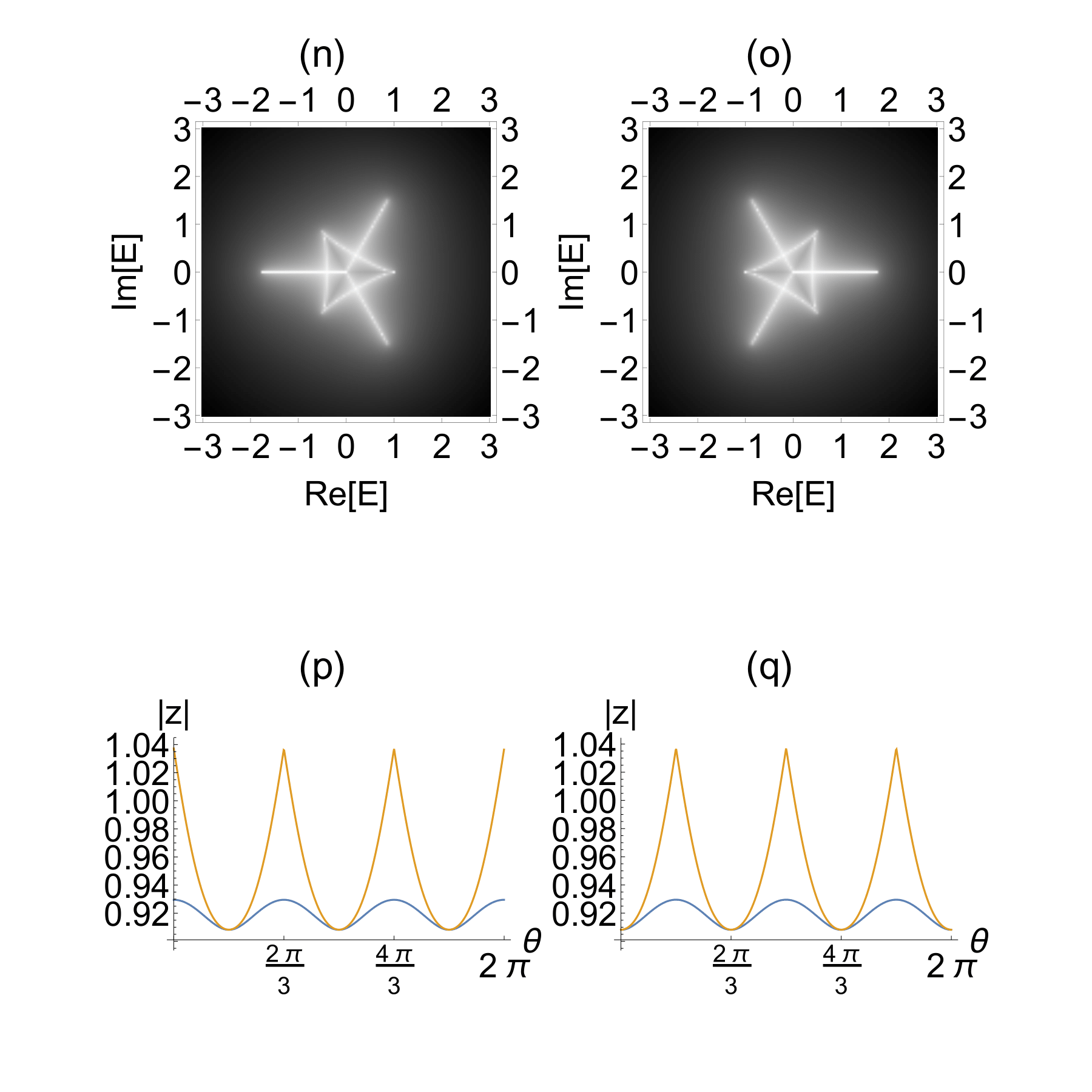}}
\end{minipage}
\caption{\textbf{Example 1 (continued).} For the polynomial $P(E,z)=z^2+1/z+rEz-E^2$, with $r=-10$, $r=-1$, $r=-0.5$, $r=0$, $r=0.5$, $r=1$, $r=10$ for (a-g); $r=0.6$, $r=0.8$, $r=1.0$, $r=2.0$, $r=5.0$, $r=8.0$ for (h-m); $r=-1$, $r=1$ for (p-q). We show the OBC spectra $\overline{E}$ in complex energy space \textbf{(a-g)}, as well as, the inverse skin depth profiles $|z|$ along the circular contour $E_0e^{i\theta}$ for constant $\theta=2\pi/3$ \textbf{(h-m)}. The inverse skin depth surfaces  $|z_i(E)|=|z_j(E)|$ trace out OBC eigenenergy solutions in (a-g). As we have a priori knowledge that the 3-wedge is there for all values of $r$, we choose any of the three theta values where we are guaranteed to have a band intersection. 
In both (a-g) and (h-m), we see that the triangle grows from small $r\neq 0$, up to $|r|=1$. Triangle shrinks beyond $|r|>1$, which is indicated by the blue arrow which covers the range of $r>1$ values that the shrinking triangle is still present (up to $r\sim8$). \textbf{(n-q)} The spectrum is mirrored about the imaginary energy axis when $r\rightarrow-r$. The inverse skin depth profile, with periodicity $2\pi/3$ also gains a $\pi$ phase shift. 
}
\label{fig:S7}
\end{figure*}

For the OBC spectra to be invariant under the simultaneous gauge transformation $\overline{E}\rightarrow \overline{E}e^{i\theta}$ and $z\rightarrow ze^{i\phi}$, we must have $\theta=2\pi/3=\phi$, i.e. 3-fold global symmetry. The OBC spectra $\overline{E}$ has a mirror symmetry about $\text{Re}(E)=0$ when $r\rightarrow-r$ (Fig.~\ref{fig:S7}n-q), so it is sufficient to just consider the evolution of the OBC eigenenergies with $|r|$. In the limit of small $|r|\neq0$, the branches continue to grow out radially at each vertex of the triangle,  such that the total number of branches at large $|\overline{E}|$ is $\mathcal{N}_0=6$ (equally spaced at $\pi/3$ intervals), where the effect of the perturbation $r$ is effectively negligible.

It will be of interest to obtain analytics for this curious triangle centred at the origin. We quantify the size of the triangle as the radius of the circle $E_0$ needed to circumscribe this triangle. Visually, this is almost equal to the magnitude of the perturbation $r$ when $|r|<1$. While the triangle is not perfectly straight, it is reasonable to assume an equilateral triangle ansatz. Let the vertices of this triangle be denoted 1, 2 and 3, such that their complex energies are
\begin{equation}
\overline{E}_1=E_0,\quad \overline{E}_2=E_0e^{2i\pi/3},\quad \overline{E}_3=E_0e^{-2i\pi/3}
\end{equation}
and the point 4 be on the purely vertical branch such that by geometry, its real value is $\text{Re}[\overline{E}_4]=\overline{E}_1\cos\pi/3=E_0/2$. The length of each triangle is $|E_0e^{2i\pi/3}-E_0|=|E_02i\sin(\pi/3)|=E_0\sqrt{3}$. We want to solve for $E_0$ in terms of $r$, which in turn gives us the size of the triangle. Without loss of generality, we take $E_0\in\mathbb{R}$  (otherwise rotate the coordinate system). Since all coefficients are real, the solutions of the characteristic polynomial are $z_{1,2}=a\pm ib$ and $z_3=c$:
\begin{align}
0&=z^3+1+rE_0z^2-E^2z\nonumber\\&=(z-z_1)(z-z_2)(z-z_3)\nonumber\\&=z^3-(c+2a)z^2+(2ac+b^2+a^2)z-(a^2+b^2)c\nonumber
\end{align}
The form of the roots is consistent with the GBZ condition where OBC eigenenergies $\overline{E}$ occur at $|z_i|=|z_j|$ for some $i\neq j$. In this case, the bands involved are $i=1$, $j=2$. Comparing coefficients give
$$-(a^2+b^2)c=1,\quad -(c+2a)=rE_0,\quad 2ac+b^2+a^2=-E_0^2$$
Solving the three equations for three unknowns, we obtain
$$E_0=r\frac{2a-(b^2+a^2)^2}{a^2+b^2}\frac{a^2+b^2}{1-2a(a^2+b^2)}=r\frac{2a-(b^2+a^2)^2}{1-2a(a^2+b^2)}$$

Fig.~\ref{fig:S7} shows the inverse skin depth profile intersections as the perturbation $|r|$ increases, keeping $|r|<1$ and given a constant circular contour $E_0=0.2$. In the regime $0<|r|<E_0$, the triangle is much smaller than the circle and the circle captures all $\mathcal{N}_0=6$ branches (effect of the perturbation $r$ is small for this value of $E_0$). Up till $|r|=E_0$, the circle circumscribes the triangle. When $|r|$ is increased further, the circle intersects at two OBC eigenenergies $\overline{E}$ in the vicinity of each triangle vertex, giving us a total of 9 solutions, inclusive of the 3 radial branches. Finally, when the triangle is large enough to circumscribe the circle, the circle will only capture the 3 radial branches.

The evolution of the spectra beyond $|r|>1$ is more interesting. The triangle initially grows bigger and subsequently shrinks in the regimes $0<|r|<1$ and $|r|>1$ respectively. When the triangle expands radially outward, the OBC branches that branch out from the triangle's vertices shrink. In the limit of large $|r|$, we are left with the 3-star as the dominant OBC solutions. This is manifested in the inverse decay length surface intersections (Fig.~\ref{fig:S7}k-m) where they now occur only at a single OBC eigenenergy value (instead of a range of $\overline{E}$ values, typical of an outward growing OBC branch) when $|r|\geq1$ for fixed $\theta=0,2\pi/3,4\pi/3$.

Above, the qualitative shape of the spectrum is largely unchanged as $r$ varies. This is because $\mathcal{N}_S=\mathcal{N}_L$ and $\gcd(g,f)=1$

\subsubsection{Example 2: $G(E)=E^2$, $F(E)=E^3$}

\begin{figure*}
\begin{minipage}{\linewidth}
\subfloat{\includegraphics[width=\linewidth]{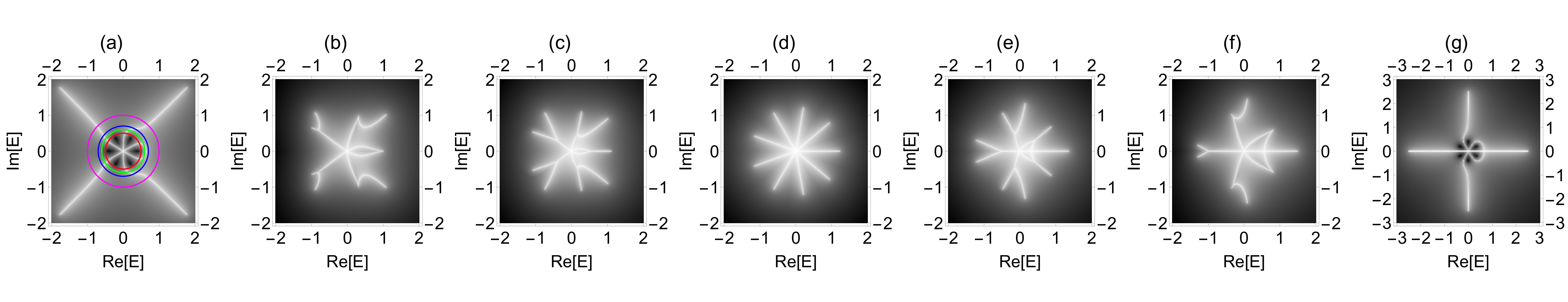}}\\
\subfloat{\includegraphics[width=\linewidth]{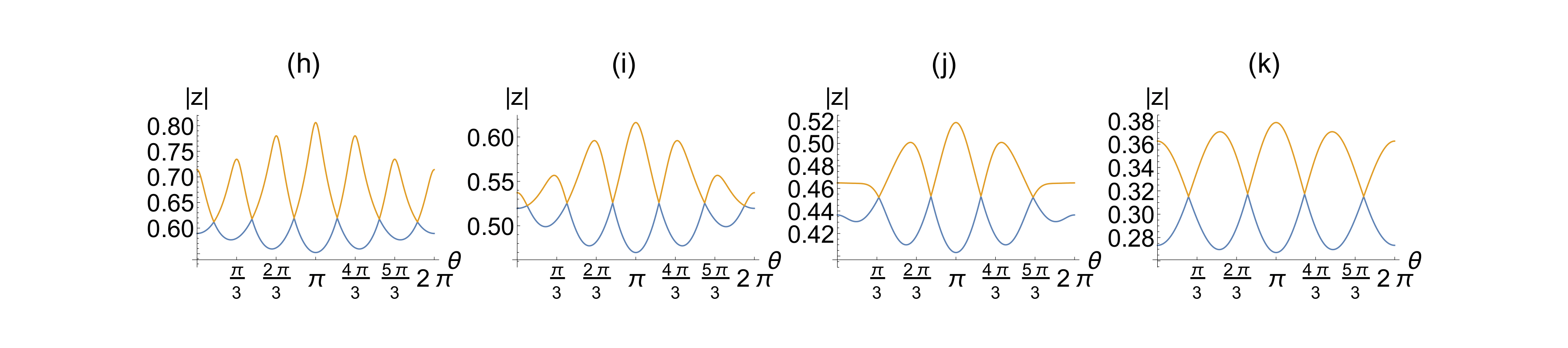}}
\end{minipage}
\caption{\textbf{Example 2.} For the polynomial $P(E,z)=z^2+1/z+rE^2z-E^3$, with $r=-10$, $r=-1$, $r=-0.5$, $r=0$, $r=0.5$, $r=1$, $r=10$ for (a-g); $r=-10$ for (h-k). \textbf{(a-g)} The evolution of OBC spectra $\overline{E}$ in complex energy space with $r$. \textbf{(h-k)} To study the OBC loops, we study the inverse skin depth profiles $|z|$ along the circular contour $E_0e^{i\theta}$ for constant $E_0$ and $r=-10$. The loop closure occurs between $\text{Re}(E)=0.6$ to $\text{Re}(E)=0.7$. By varying $E_0$, we can count the band intersections $n$. (h) $E_0=0.5$, $n=6$ (red circle); (i) $E_0=0.6$, $n=6$ (green circle); (j) $E_0=0.7$ (blue circle), $n=4$; (k) $E_0=1$, $n=4$ (magenta circle).}
\label{fig:S8}
\end{figure*}

Fig.~\ref{fig:S8} shows the OBC spectra for $G(E)=E^2$ and $F(E)=E^3$ for various $r$ values. The spectra exhibit a greater level of complexity compared to the preceding example, hence a detailed analysis is difficult. When $r=0$, we have an expected $\mathcal{N}_0=3f=9$-wedge star. When $r\neq 0$ is small, some of the radial branches fuse in the small $|\overline{E}|$ region such that there are only $\mathcal{N}_S=6$ wedges in this neighbourhood. As $\mathcal{N}_0>\mathcal{N}_S$, we have the branches to bifurcate as they branch radially outward. When $r$ is large enough, the branches at the large $|\overline{E}|$ limit will change from $\mathcal{N}_0$ to $\mathcal{N}_L$. But $\mathcal{N}_L=4<\mathcal{N}_S=6$, then by continuity, the OBC eigenenergies $\overline{E}$ will loop into itself. Again, we demonstrate this using the inverse surface depth profiles in Fig.~\ref{fig:S8}. The number of band intersections decrease from 6 at small $|E_0|$, up till $|E_0|\sim0.62$ where the loop occurs, to 4 at large $|E_0|$.

\subsubsection{Example 3: $G(E)=E^3$, $F(E)=E^4$}
\begin{figure*}
\centering
\includegraphics[width=\linewidth]{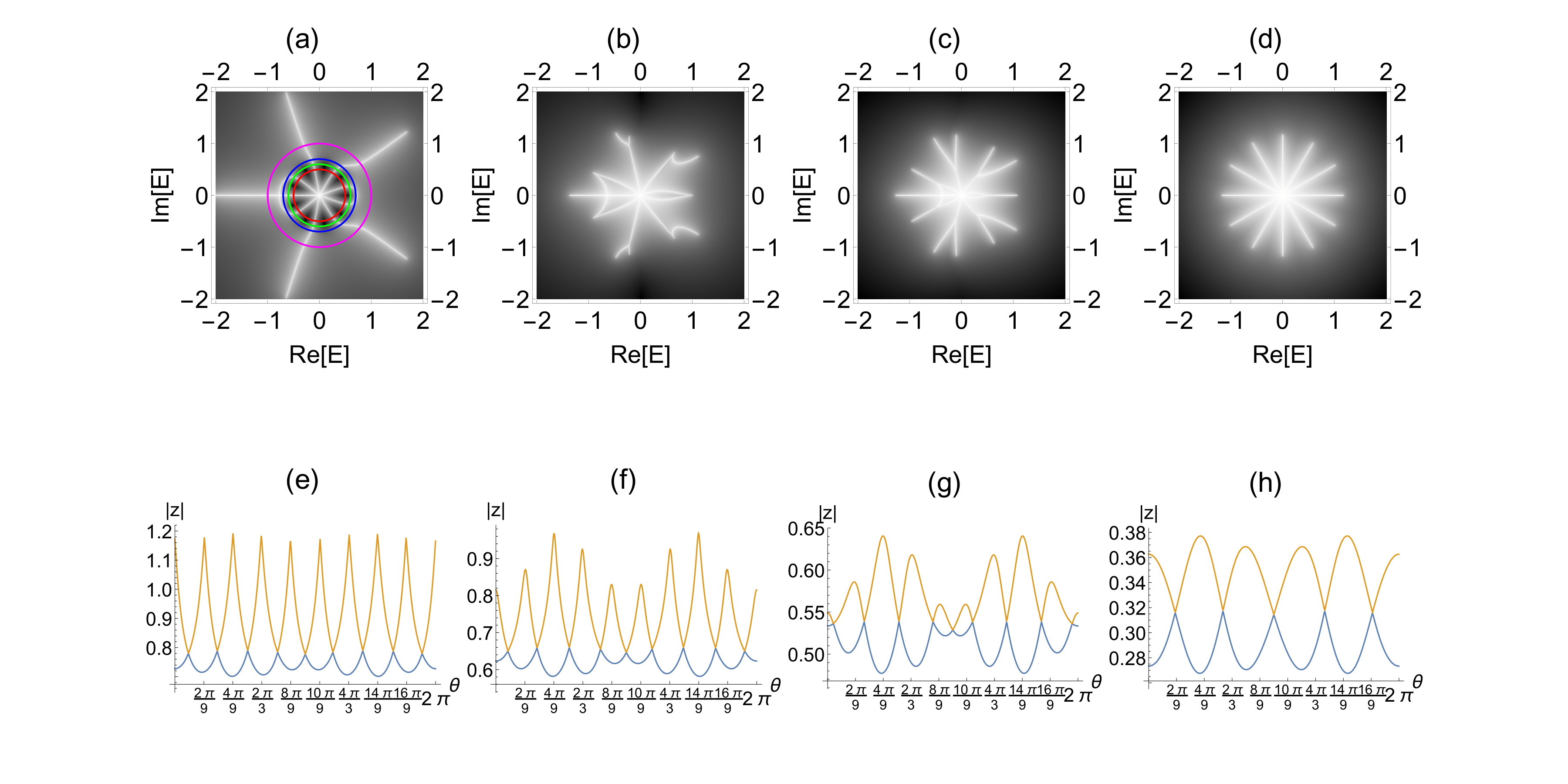}
\caption{\textbf{Example 3.} For the polynomial $P(E,z)=z^2+1/z+rE^3z-E^4$, with $r=-10$, $r=-1$, $r=-0.5$, $r=0$ for (a-d); $r=-10$ for (e-h). \textbf{(a-d)} Evolution of OBC spectra from an insect-like to flower-like shape. with $r$.
\textbf{(e-h)} To study the two closed loops in the OBC spectrum for $r=-10$, we study the inverse skin depth profiles $|z|$ along the circular contour $E_0e^{i\theta}$ for constant $E_0$. The right loop closes up at $E_0=0.8$ and only a point is captured. Beyond that, at $E_0=1$, the is not captured. Here, the left loop has also looped up and is consequentially not captured as well. By varying $E_0$, we can count the surface intersections $n$. (e) $E_0=0.5$, $n=9$ (red circle); (f) $E_0=0.6$, $n=9$ (green circle); (g) $E_0=0.7$, $n=8$ (blue circle); (h) $E_0=1$, $n=5$ (magenta circle).
}
\label{fig:S9}
\end{figure*}

Fig.~\ref{fig:S9} shows the OBC spectrum for $G(E)=E^3$ and $F(E)=E^4$ for various $r$ values. As $g$ is odd and $f$ is even, the spectrum transforms like $E\rightarrow -E$ when $r\rightarrow -r$, i.e. mirror image. When $r=0$, we have an expected $\mathcal{N}_0=3f=12$-wedge star. When $r\neq 0$ is small, some of the radial branches fuse in the small $|E|$ region such that there are only $\mathcal{N}_S=6$ wedges in this neighbourhood. Again, as $\mathcal{N}_0>\mathcal{N}_S$, we have the branches to bifurcate as they branch radially outward. As $|r|$ increases, one of the two branched out arms at each bifurcation point dominate the `competition' and survives while the other shrinks. As $\mathcal{N}_S=9>\mathcal{N}_L=5$, we expect again the OBC eigenenergies to loop. This time, there are only $(\mathcal{N}_S-\mathcal{N}_L)/2=2$ closed loops. At large $|r|$, the dominant OBC modes will then be the superposition of both loops and an outer 5-wedge star. In the inverse skin depth profile (Fig.~\ref{fig:S9}), the number of surface intersections decrease from 9 to 8 (when $E_0$ is big enough to touch the tip of the smaller closed loop) and then to 5 (when $E_0$ misses both closed loops, leaving the dominant 5-star). 

\newpage
\subsubsection{Example 4: $G(E)=E$, $F(E)=E^3$}
\begin{figure*}
\centering
\includegraphics[width=\linewidth]{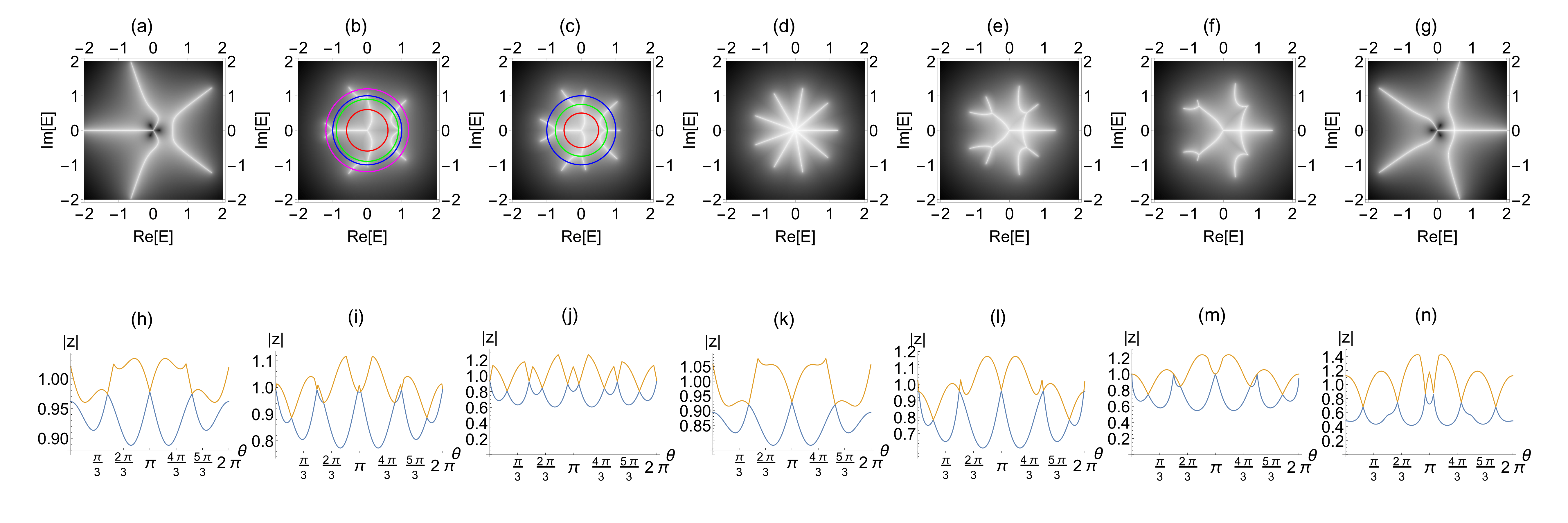}
\caption{\textbf{Example 4.} For the polynomial $P(E,z)=z^2+1/z+rEz -E^3$, we show the OBC spectra $\overline{E}$ in complex energy space (a-g) and the inverse skin depth profiles $|z|$ along the circular contour $E_0 e^{i\theta}$ for constant $r=-0.5$ (h-j) and $r=-1$ (k-n). When two inverse skin depth surfaces touch, i.e. $|z_i(E)|=|z_j(E)|$, these band intersections (the number of such intersections is $n$) in (h-n) trace out OBC eigenenergy solutions in (a-g). \textbf{(a-g)} Evolution of OBC eigenenergies with $r$. \textbf{(h-n)} In particular, we are interested in the OBC eigenenergies for $r<0$ where an additional branch is found isolated from the rest of the spectrum. $E_0<0.65$ cannot capture this segregated branch and this `skin gap' shows. (h) $E_0=0.5$ (red circle), (i) $E_0=0.75$ (green circle), (j) $E_0=1$ (blue circle); (k) $E_0=0.6$ (red circle), (l) $E_0=0.9$ (green circle), (m) $E_0=1.0$ (blue circle), (n) $E_0=1.2$ (magenta circle).
}
\label{fig:S10}
\end{figure*}

Fig.~\ref{fig:S10} shows the OBC spectrum for $G(E)=E$ and $F(E)=E^3$ for various $r$ values. The evolution of the spectrum with $r$ is more complicated, demonstrating bifurcation and subsequently an additional branch appears when $r<0$ is sufficiently large. When $r=0$, we have an expected $\mathcal{N}_0=3f=9$-wedge star.  When $r\neq 0$ is small, there is only $\mathcal{N}_S=3$ wedges in the small $|E|$ neighbourhood. Here, there is no relation between the spectrum with small $r$ and $-r$, and are perturbed differently when $r$ is switched on. For $r=0.5$, 6 of the radial branches fuse, giving 3 near the centre, and by continuity arguments, we have $\mathcal{N}_0=9$ branches in the large $|E|$ neighbourhood for small $r>0$.  For $r=-0.5$, the OBC eigenenergies rearrange themselves such that we have a segregated branch on the right. From Fig.~\ref{fig:S10}(b-d) ($r=-0.5$), we can see that for $E_0<0.65$, we cannot capture the additional branch (along $\theta=0=2\pi$) on the right, demonstrating the `skin gap'. A large enough $E_0$ can capture all $\mathcal{N}_0=12$ wedges at the terminations (same as $r=0$). This additional branch is consistent with the fact that $\mathcal{N}_L=5>\mathcal{N}_S=3$. At a more negative $r$ value $r=-1$ (Fig.~\ref{fig:S10}(e-h)), we see that the number of surface intersections captured increases from $\mathcal{N}_S=3$, to 7 (captures the right side branch), followed by 9, and finally 6 $\neq\mathcal{N}_0$ since the branches shrink due to competition again.

\begin{figure*}
\begin{minipage}{\linewidth}
\subfloat{\includegraphics[width=0.5\linewidth]{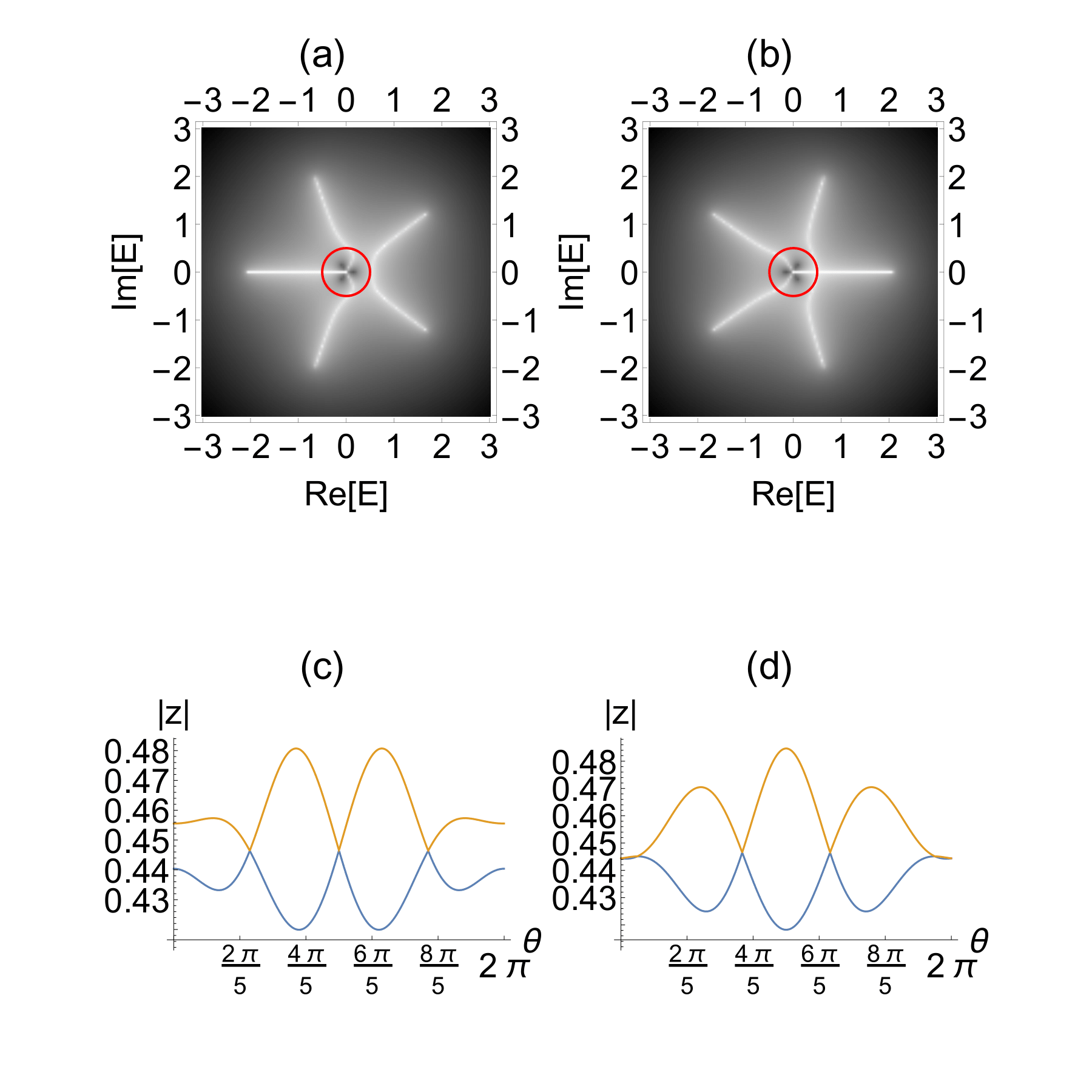}}
\subfloat{\includegraphics[width=0.5\linewidth]{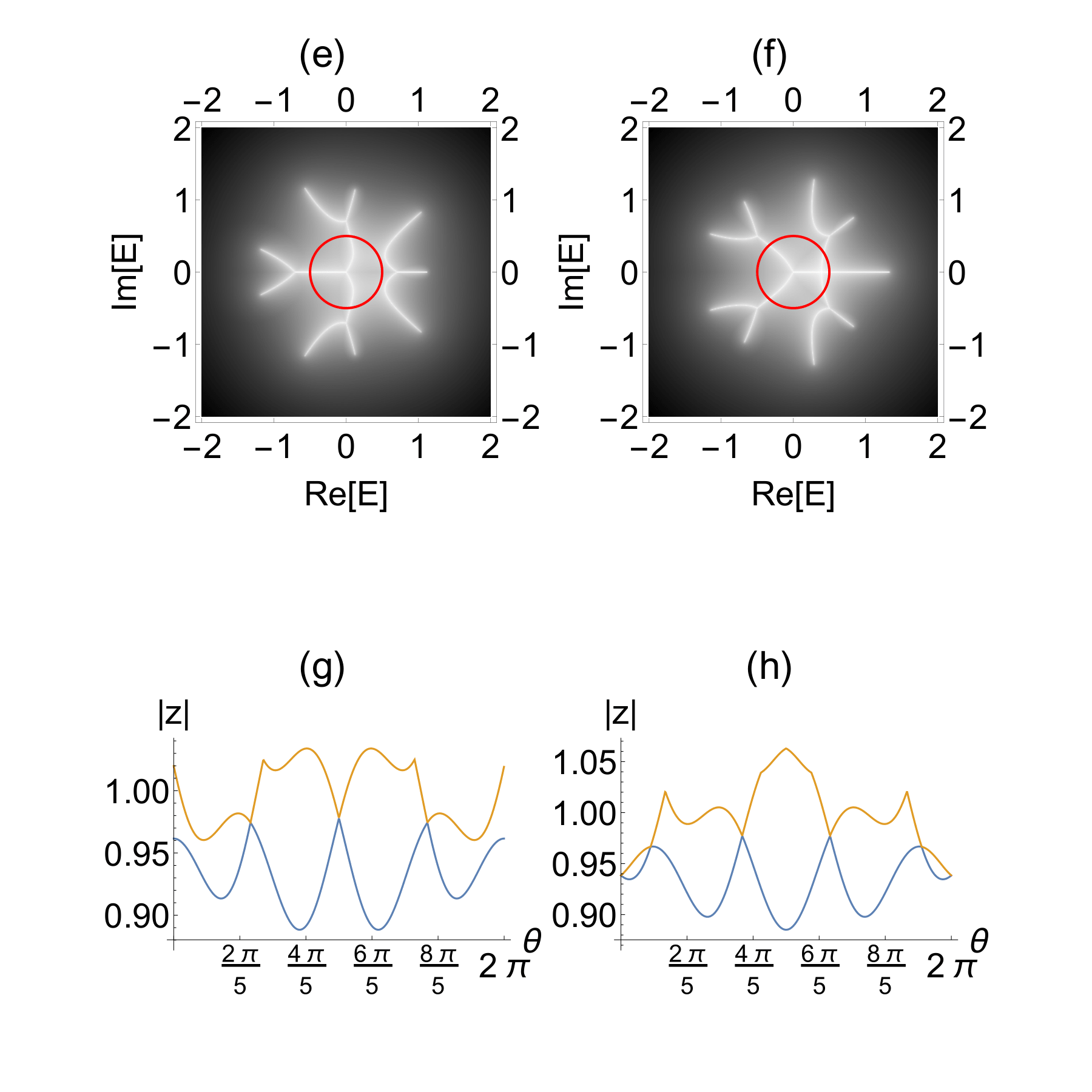}}
\end{minipage}
\caption{\textbf{Misleading symmetry, illustrated with Example 4.} For the polynomial $P(E,z)=z^2+1/z+rEz -E^3$, we show the skin spectra $\overline{E}$ in complex energy space (a-g) and the inverse skin depth profiles $|z|$ along the circular contour $E_0 e^{i\theta}$  (in red) for constant $r$ and $E_0=0.5$. When two bands touch $|z_i(E)|=|z_j(E)|$, these band intersections (the number of such intersections is $n$) trace out OBC eigenenergy solutions. \textbf{(a-d)} Comparing $r=-10,10$, the OBC eigenenergies (a-b) appear related by reflection symmetry $E\rightarrow -E$ as $r\rightarrow-r$. But this the inverse skin depth profiles (c-d) are not related by $\pi$ phase shift. \textbf{(e-h)} Comparing this with $r=-0.5,0.5$, the OBC eigenenergies (e-f) appear completely different. Their respective inverse skin depth profiles (g-h) are shown for completeness.
}
\label{fig:S11}
\end{figure*}

Finally, we like to highlight in Fig.~\ref{fig:S11} the `misleading symmetry' for large $r$ and large $|E|$: a quick inspection might lead one to conclude $r\rightarrow -r$ results in $E\rightarrow -E$. This is approximately true since $P(E,z)\approx z^2+(1/z)-r E z$. But for small $|r|$, this is completely not true.

\begin{figure*}
\centering
\includegraphics[width=\linewidth]{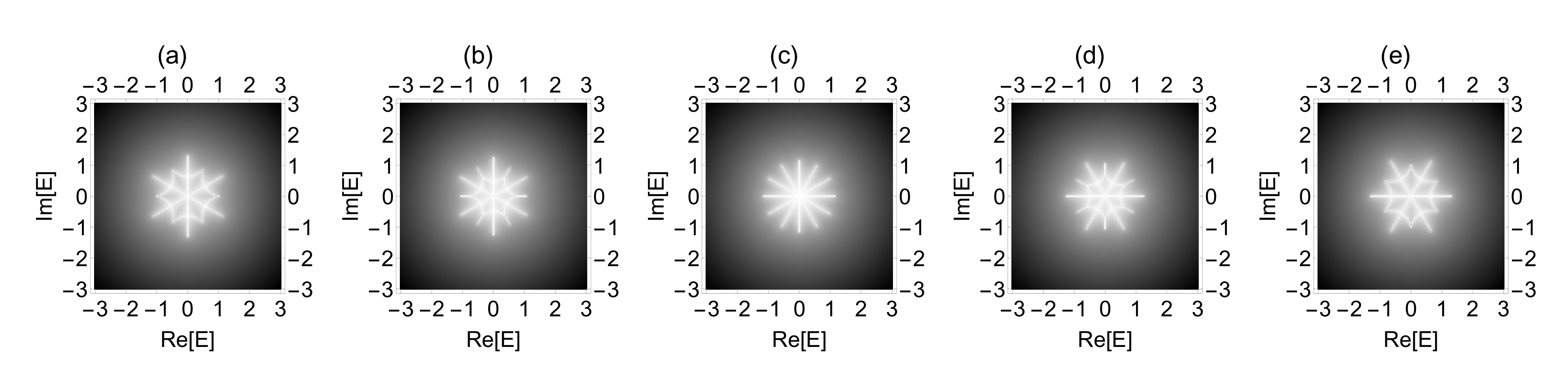}
\caption{\textbf{Example 5: Absence of branch bifurcation.} For the polynomial $P(E,z)=z^2+1/z+rE^2z -E^4$, we show how the OBC spectra evolves with $r$. No bifurcation or looping occurs since $\mathcal{N}_S=\mathcal{N}_L$. (a) $r=-1$, (b) $r=-0.5$, (c) $r=0$, (d) $r=0.5$, (e) $r=1$.
}
\label{fig:S12}
\end{figure*}
\subsubsection{Example 5: $G(E)=E^2$, $F(E)=E^4$}
In the last example (\ref{fig:S11}), we see again $\mathcal{N}_S=\mathcal{N}_L$ (so no bifurcation or looping occurs), but this time the evolution of the OBC spectrum is more interesting than the first case. The two are related via a conformal map $E\rightarrow E^2$.

\newpage
\section{Adjacency matrices of various spectral graphs}
Here we list down the adjacency matrices of the spectral graphs presented in Table I of the main text. The adjacency matrices are invariants of the graph topology, up to permutations of the graph vertices.
\setcounter{MaxMatrixCols}{25}
\begin{itemize}
\item (i) $P(E,z)=z^2+1/z+rEz-E^2$: $\begin{pmatrix}
0 &1 &1 &1 &1 &0 &0\\ 
1 &0 &1 &1 &0 &1 &0\\ 
1 &1 &0 &1 &0 &0 &1\\ 
1 &1 &1 &0 &0 &0 &0\\ 
1 &0 &0 &0 &0 &0 &0\\ 
0 &1 &0 &0 &0 &0 &0\\ 
0 &0 &1 &0 &0 &0 &0
\end{pmatrix}$
\item (ii) $P(E,z)=z^2+1/z+rE^2z-E^4$: $\begin{pmatrix}
0 &1 &0 &0 &0 &1 &1 &1 &0 &0 &0 &0 &0\\ 
1 &0 &1 &0 &0 &0 &1 &0 &1 &0 &0 &0 &0\\ 
0 &1 &0 &1 &0 &0 &1 &0 &0 &1 &0 &0 &0\\ 
0 &0 &1 &0 &1 &0 &1 &0 &0 &0 &1 &0 &0\\ 
0 &0 &0 &1 &0 &1 &1 &0 &0 &0 &0 &1 &0\\ 
1 &0 &0 &0 &1 &0 &1 &0 &0 &0 &0 &0 &1\\ 
1 &1 &1 &1 &1 &1 &0 &0 &0 &0 &0 &0 &0\\ 
1 &0 &0 &0 &0 &0 &0 &0 &0 &0 &0 &0 &0\\ 
0 &1 &0 &0 &0 &0 &0 &0 &0 &0 &0 &0 &0\\ 
0 &0 &1 &0 &0 &0 &0 &0 &0 &0 &0 &0 &0\\ 
0 &0 &0 &1 &0 &0 &0 &0 &0 &0 &0 &0 &0\\ 
0 &0 &0 &0 &1 &0 &0 &0 &0 &0 &0 &0 &0\\ 
0 &0 &0 &0 &0 &1 &0 &0 &0 &0 &0 &0 &0 
\end{pmatrix}$
\item (iii) $P(E,z)=z^2+1/z+rEz-E^3$: $\begin{pmatrix}
0 &0 &0 &1 &0 &0 \\ 
0 &0 &0 &1 &0 &0 \\ 
0 &0 &0 &1 &0 &0 \\ 
1 &1 &1 &0 &0 &0 \\ 
0 &0 &0 &0 &0 &1 \\ 
0 &0 &0 &0 &1 &0 \\ 
\end{pmatrix}$
\item (iv) $P(E,z)=z^3+1/z^2+rEz-E^3$: $\begin{pmatrix}
0 &0 &0 &0 &0 &1 &1 &1 &0 &0 &0 &0 &0\\ 
0 &0 &0 &0 &0 &1 &0 &0 &1 &1 &0 &0 &0\\ 
0 &0 &0 &1 &0 &1 &0 &0 &0 &0 &1 &0 &0\\ 
0 &0 &1 &0 &1 &1 &0 &0 &0 &0 &0 &1 &0\\ 
0 &0 &0 &1 &0 &1 &0 &0 &0 &0 &0 &0 &1\\ 
1 &1 &1 &1 &1 &0 &0 &0 &0 &0 &0 &0 &0\\ 
1 &0 &0 &0 &0 &0 &0 &0 &0 &0 &0 &0 &0\\ 
1 &0 &0 &0 &0 &0 &0 &0 &0 &0 &0 &0 &0\\ 
0 &1 &0 &0 &0 &0 &0 &0 &0 &0 &0 &0 &0\\ 
0 &1 &0 &0 &0 &0 &0 &0 &0 &0 &0 &0 &0\\ 
0 &0 &1 &0 &0 &0 &0 &0 &0 &0 &0 &0 &0\\ 
0 &0 &0 &1 &0 &0 &0 &0 &0 &0 &0 &0 &0\\ 
0 &0 &0 &0 &1 &0 &0 &0 &0 &0 &0 &0 &0 
\end{pmatrix}$
\item (v) $P(E,z)=z^2+1/z+rE^2z-E^3$: $\begin{pmatrix}
0 &3 &1 &1 &1 &0 \\
3 &0 &0 &0 &0 &1 \\
1 &0 &0 &0 &0 &0 \\
1 &0 &0 &0 &0 &0 \\
1 &0 &0 &0 &0 &0 \\
0 &1 &0 &0 &0 &0
\end{pmatrix}$
\item (vi) $P(E,z)=z^2+1/z+rE^3z-E^4$: $\begin{pmatrix}
1 &3 &0 &1 &1 &1 &1 \\
3 &0 &1 &0 &0 &0 &0 \\
0 &1 &0 &0 &0 &0 &0 \\
1 &0 &0 &0 &0 &0 &0 \\
1 &0 &0 &0 &0 &0 &0 \\
1 &0 &0 &0 &0 &0 &0 \\
1 &0 &0 &0 &0 &0 &0 
\end{pmatrix}$
\end{itemize}

\section{Compendium of various exotic spectra}
As we have seen, exotic OBC spectral shapes can manifest in various ways - the individual branches may branch out to two or more branches each, each of these subbranches may grow or shrink accordingly, or even form closed loops. From the various examples discussed, we can understand them by counting the number of branches as $|E|$ increases from small to large values, for a given perturbation strength $r$. As previously mentioned, there should in principle be $\mathcal{N}_S=3g$ and $\mathcal{N}_L=2f-g$ branches at sufficiently small and large $|E|$ respectively. The general rules are thus:
\begin{enumerate}
\item When additional disjointed/isolated branches occur, $\mathcal{N}_S<\mathcal{N}_L$ but this condition alone does not guarantee the existence of isolated branches. 
\item Generally, the branches do not bifurcate if $\mathcal{N}_L=\mathcal{N}_S$.  One example is shown in Fig.~\ref{fig:S12}. An exception might occur if the `pattern' at large $|\overline{E}|$ is identical to that at small $|\overline{E}|$ but displaced by a small rotation angle. Typically, branches will bifurcate to ensure a smooth interpolation  between OBC modes at small $|\overline{E}|$ and large $|\overline{E}|$.
\item Sometimes, branches may develop into loops. To determine the total number of loops (if any), one may quote the renowned topological relationship for any simplicial complex on a 2D plane:
\begin{equation}
    \mathcal{V}+\mathcal{F}-\mathcal{E}=1+\mathcal{C}
\end{equation}
where $\mathcal{V}$ is the number of vertices (including the ends of each branch), $\mathcal{E}$ is the number of edges, $\mathcal{C}$ is the number of disconnected parts of the graph and $\mathcal{F}$ is the number of disconnected regions in the plane. Since each loop partitions a region into two, the total number of distinct loops is
\begin{equation}
    \mathcal{N}_\ell=\mathcal{C}+\mathcal{E}-\mathcal{V}
\end{equation}
\end{enumerate}
In the following, we present a plethora of sophisticated examples, beyond what has been presented. For each case, note the crossover from large $|r|$ patterns to the $r=0$ pattern as $r$ evolves - the interpolation between the branching behaviors of these two distinct cases is what leads to even more intricate spectral graphs. Importantly, the OBC spectra can be deduced from the above generic rules and symmetry rules discussed earlier. 


\newpage
\begin{figure*}
\centering
\includegraphics[width=0.95\linewidth]{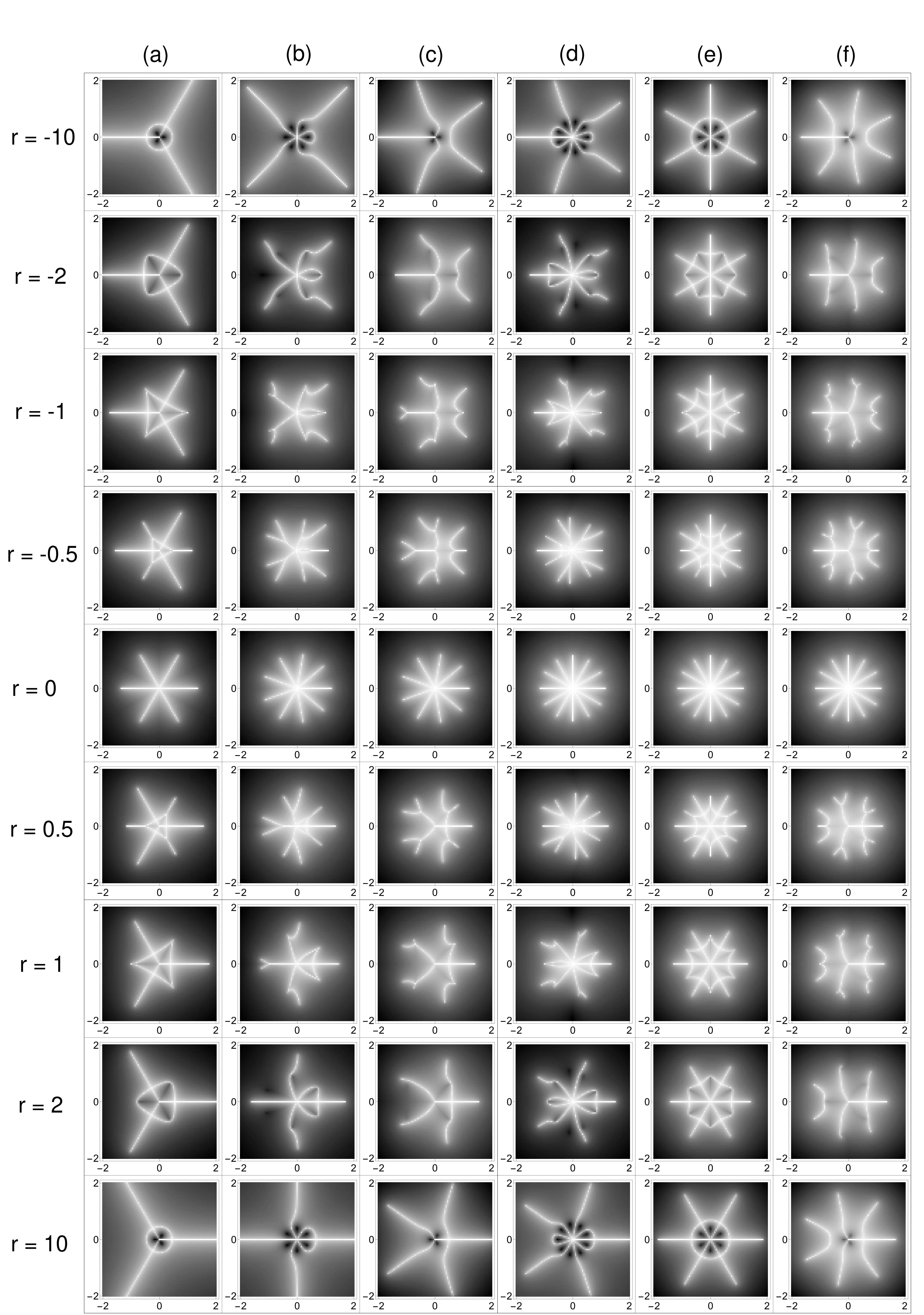}
\caption{Complex energy plots of the OBC spectra, for the polynomial $z^2 + 1/z + r E^g z - E^f$; (a) $g=1$, $f=2$, (b) $g=2$, $f=3$, (c) $g=1$, $f=3$ (d) $g=3$, $f=4$, (e) $g=2$, $f=4$, (f) $g=1$, $f=4$.
}
\label{fig:giganticfig1}
\end{figure*}
\newpage
\begin{figure*}
\centering
\includegraphics[width=0.95\linewidth]{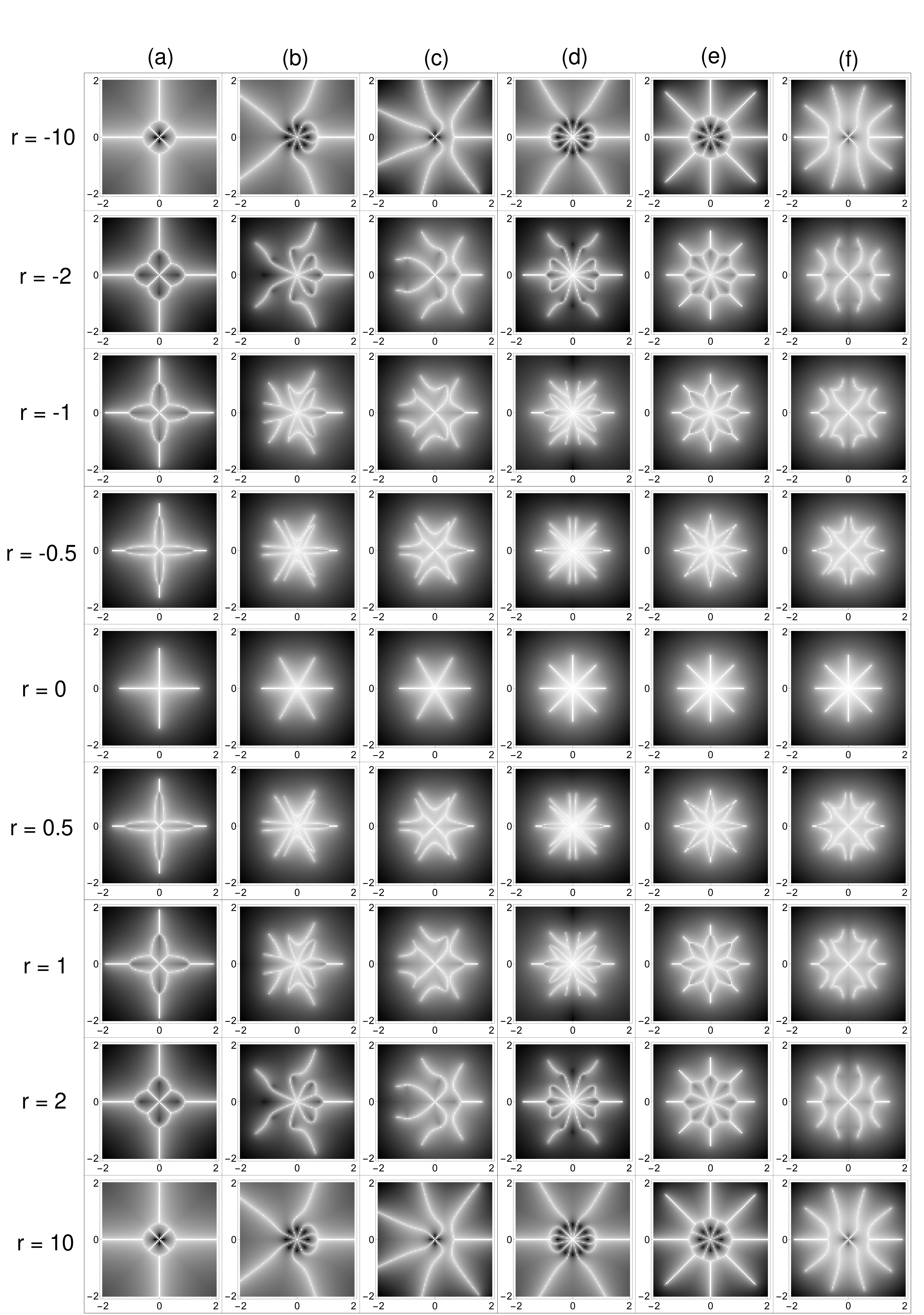}
\caption{Complex energy plots of the OBC spectra, for the polynomial $z^2 + 1/z^2 + r E^g z - E^f$; (a) $g=1$, $f=2$, (b) $g=2$, $f=3$, (c) $g=1$, $f=3$ (d) $g=3$, $f=4$, (e) $g=2$, $f=4$, (f) $g=1$, $f=4$.
}
\label{fig:giganticfig2}
\end{figure*}
\newpage
\begin{figure*}
\centering
\includegraphics[width=0.95\linewidth]{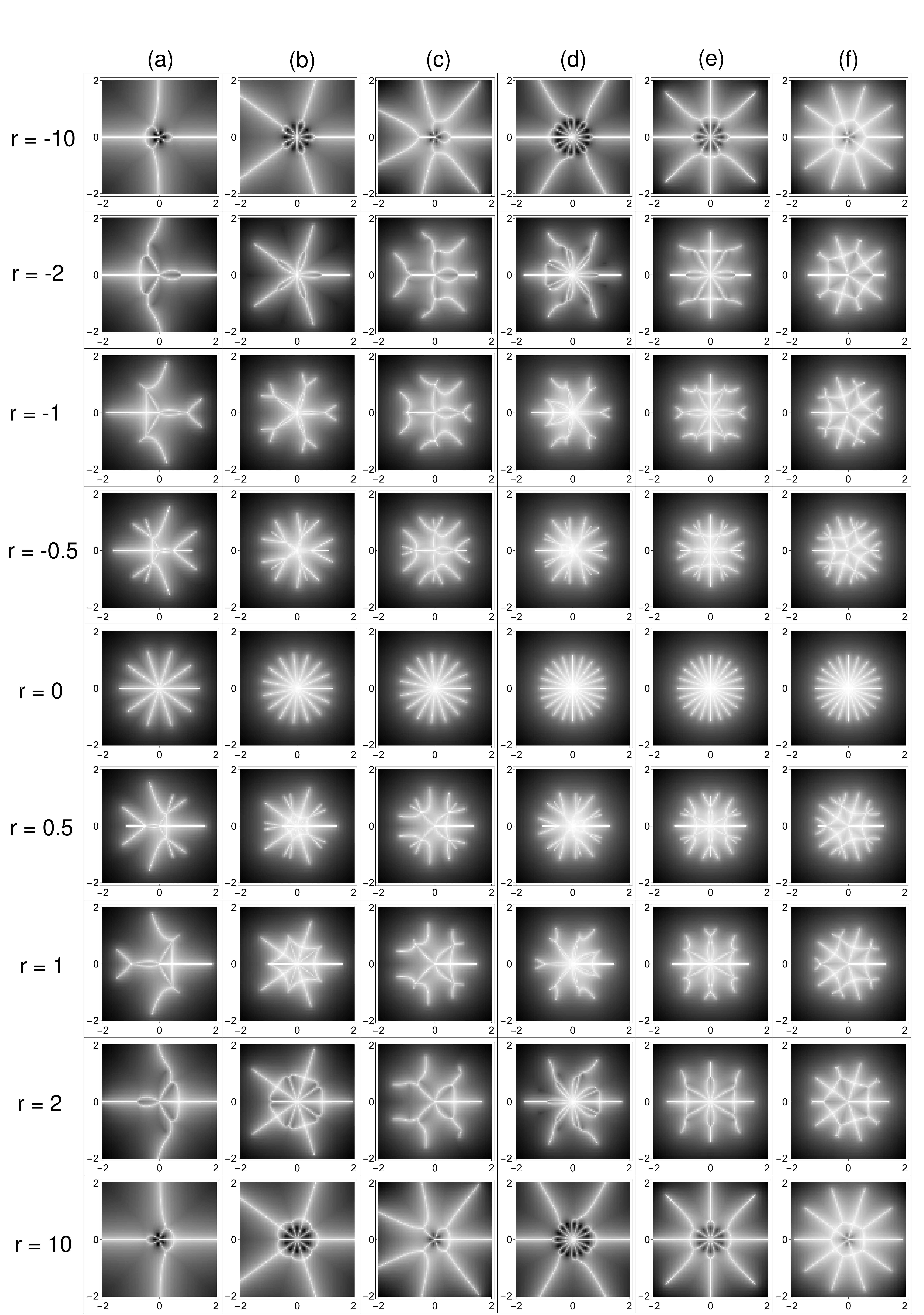}
\caption{Complex energy plots of the OBC spectra, for the polynomial $z^3 + 1/z^2 + r E^g z - E^f$; (a) $g=1$, $f=2$, (b) $g=2$, $f=3$, (c) $g=1$, $f=3$ (d) $g=3$, $f=4$, (e) $g=2$, $f=4$, (f) $g=1$, $f=4$.
}
\label{fig:giganticfig3}
\end{figure*}
\newpage
\begin{figure*}
\centering
\includegraphics[width=0.95\linewidth]{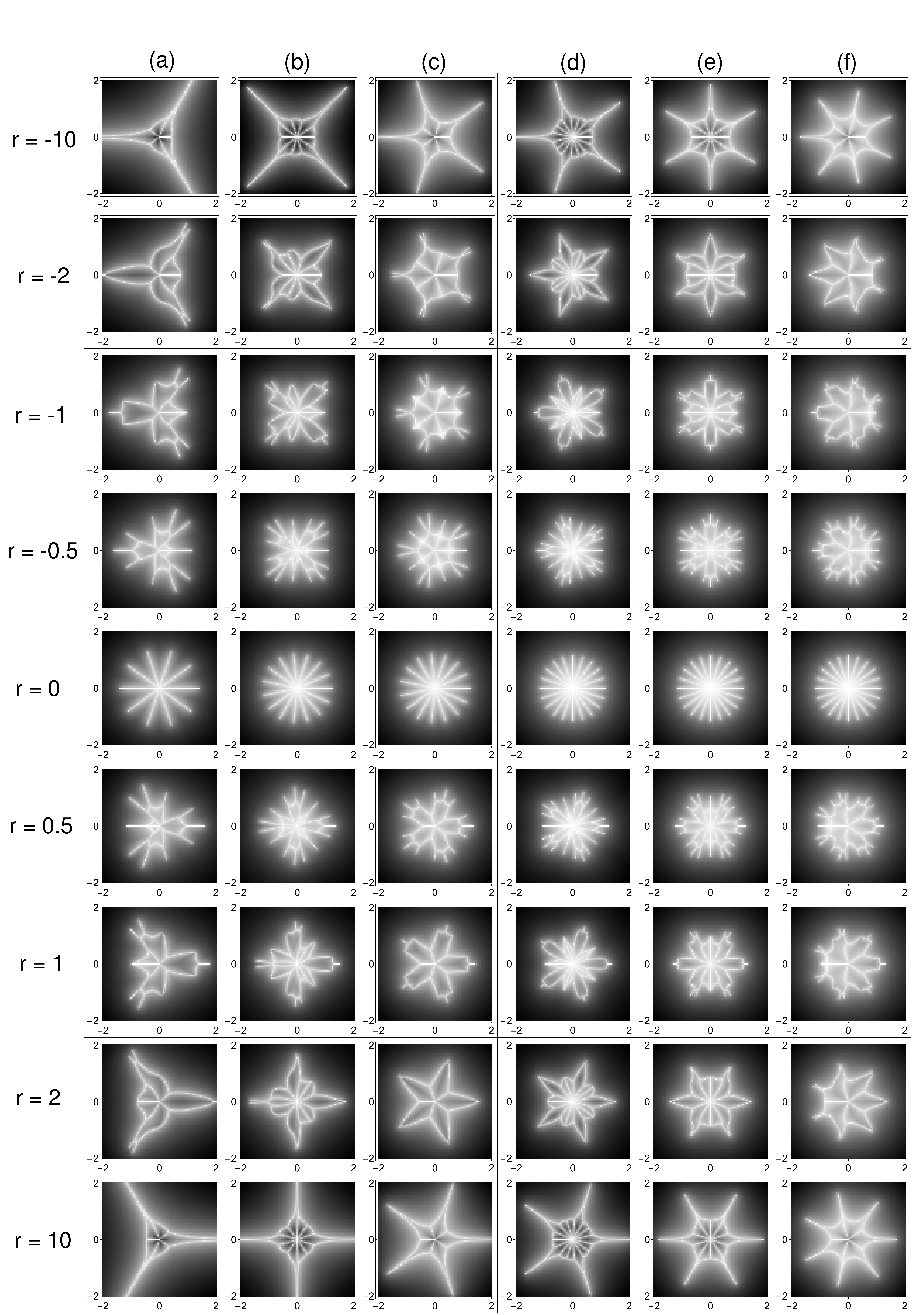}
\caption{Complex energy plots of the OBC spectra, for the polynomial $z^3 + 1/z^2 + r E^g z^2 - E^f$; (a) $g=1$, $f=2$, (b) $g=2$, $f=3$, (c) $g=1$, $f=3$ (d) $g=3$, $f=4$, (e) $g=2$, $f=4$, (f) $g=1$, $f=4$.
}
\label{fig:giganticfig4}
\end{figure*}
\newpage
\begin{figure*}
\centering
\includegraphics[width=0.95\linewidth]{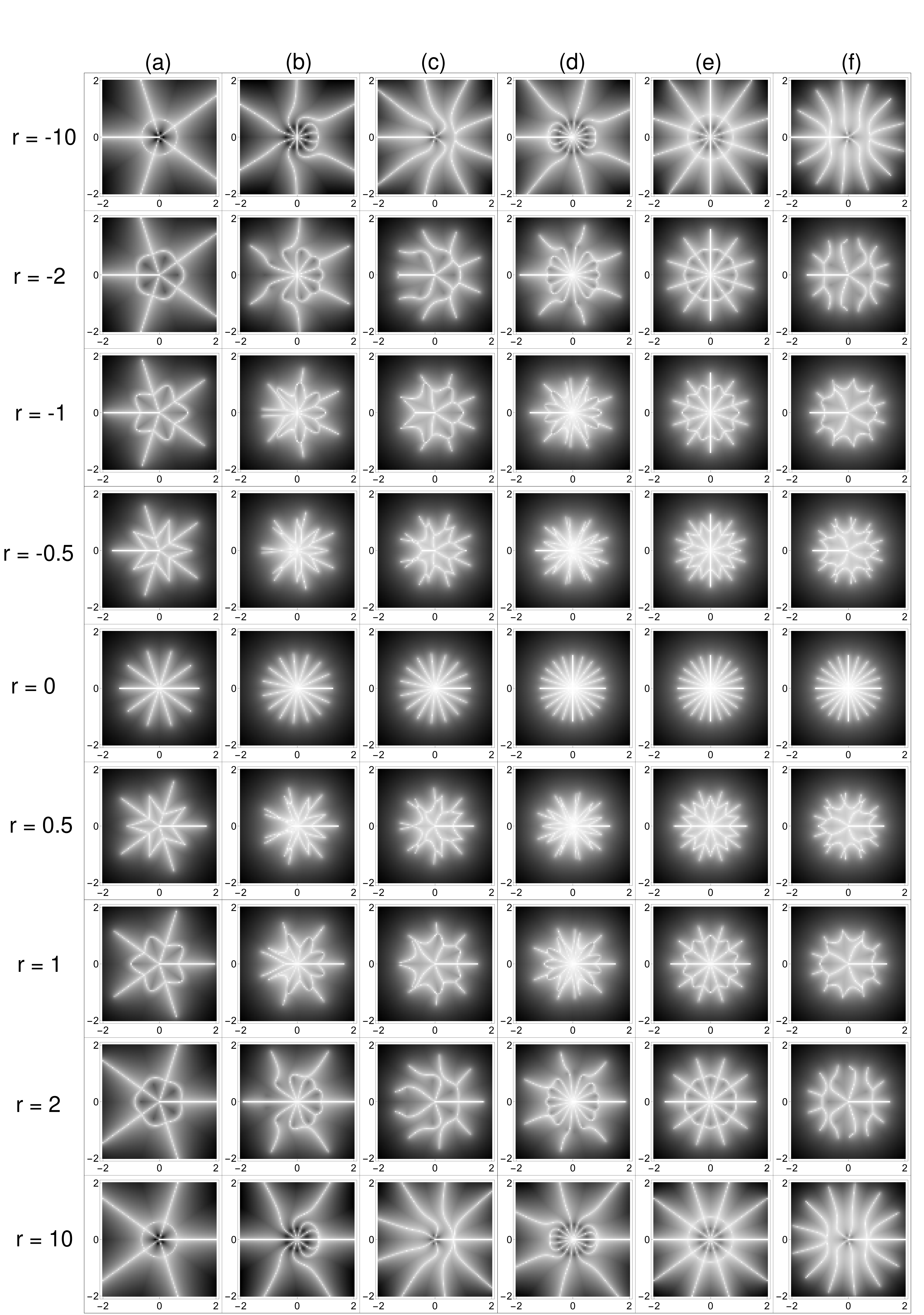}
\caption{Complex energy plots of the OBC spectra, for the polynomial $z^2 + 1/z^2 + r E^g/z - E^f$; (a) $g=1$, $f=2$, (b) $g=2$, $f=3$, (c) $g=1$, $f=3$ (d) $g=3$, $f=4$, (e) $g=2$, $f=4$, (f) $g=1$, $f=4$.
}
\label{fig:giganticfig5}
\end{figure*}
\newpage
\begin{figure*}
\centering
\includegraphics[width=0.95\linewidth]{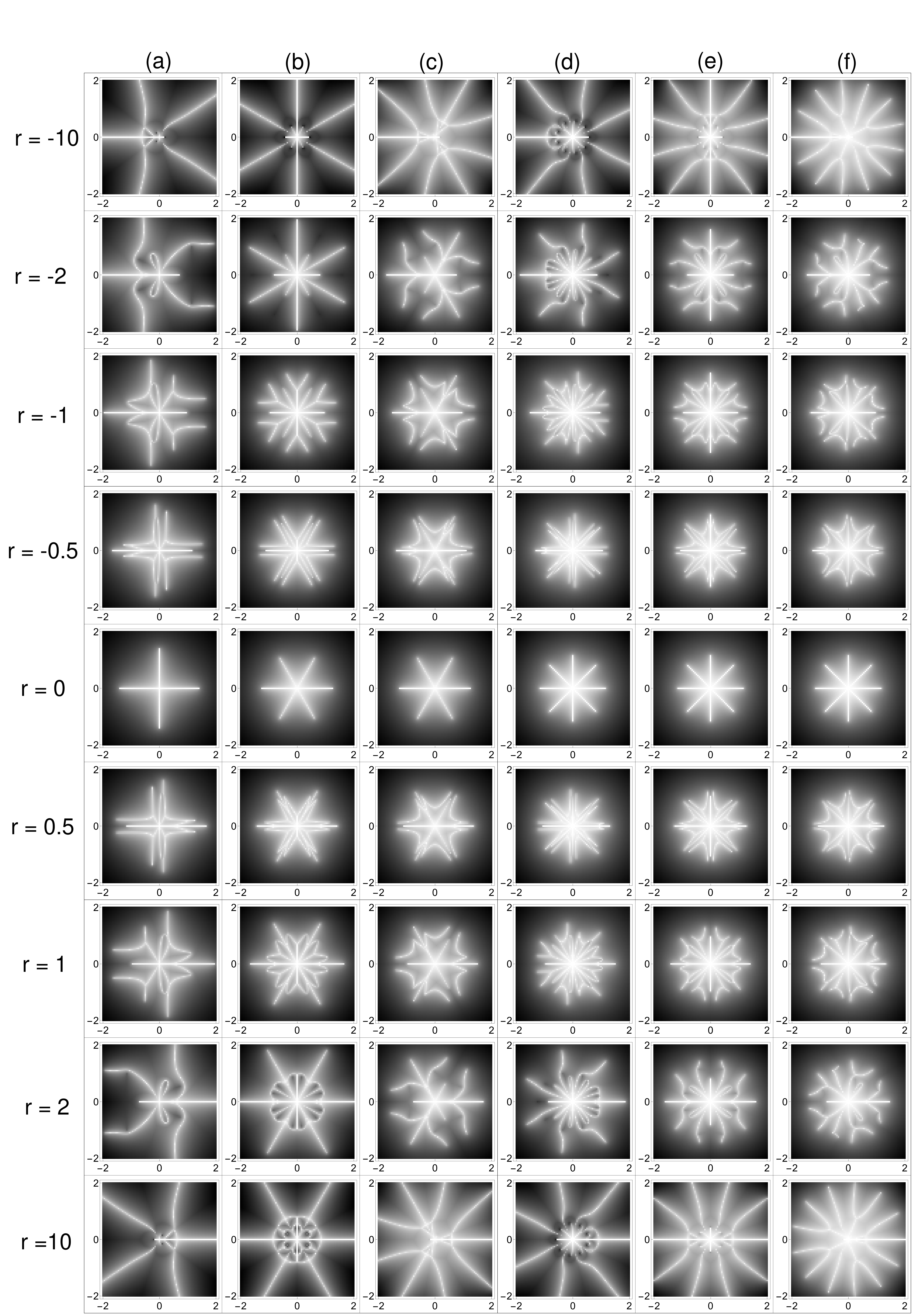}
\caption{Complex energy plots of the OBC spectra, for the polynomial $z^3 + 1/z^3 + r E^g z - E^f$; (a) $g=1$, $f=2$, (b) $g=2$, $f=3$, (c) $g=1$, $f=3$ (d) $g=3$, $f=4$, (e) $g=2$, $f=4$, (f) $g=1$, $f=4$.
}
\label{fig:giganticfig6}
\end{figure*}
\newpage
\begin{figure*}
\centering
\includegraphics[width=0.95\linewidth]{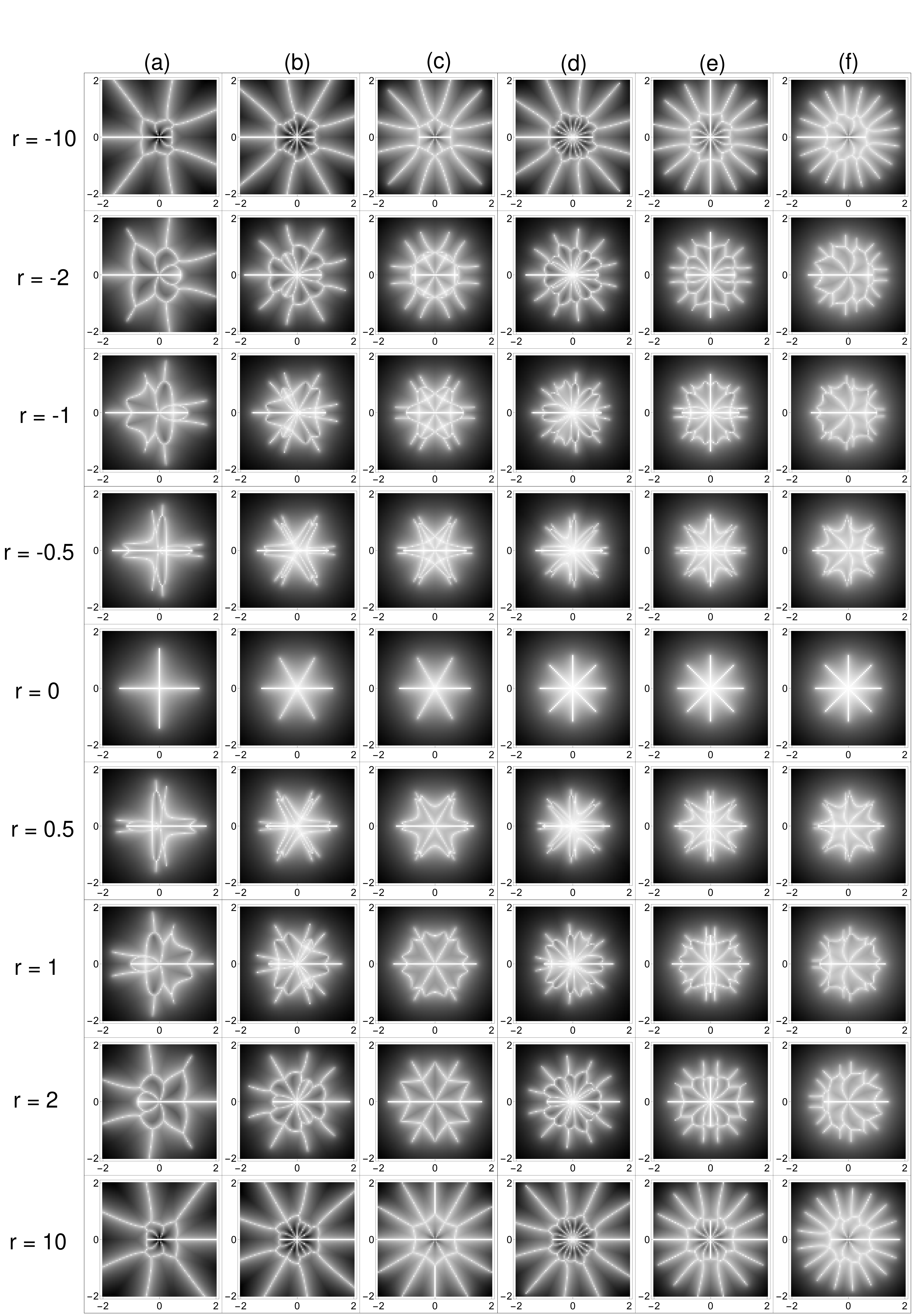}
\caption{Complex energy plots of the OBC spectra, for the polynomial $z^3 + 1/z^3 + r E^g z^2 - E^f$; (a) $g=1$, $f=2$, (b) $g=2$, $f=3$, (c) $g=1$, $f=3$ (d) $g=3$, $f=4$, (e) $g=2$, $f=4$, (f) $g=1$, $f=4$.
}
\label{fig:giganticfig7}
\end{figure*}


\end{document}